%Paper: gr-qc/9209013
%From: NARDELLI@ITNVAX.CINECA.IT
%Date: Wed, 30 SEP 92 13:23 N
%Date (revised): Thu, 5 NOV 92 10:25 N

\def\cancella#1#2{\ooalign{$\hfil#1\mkern1mu/\hfil$\crcr$#1#2$}}
\def\gam#1{\mathrel{\mathpalette\cancella{#1}}}
\def\square{\kern1pt\vbox{\hrule height 1.2pt\hbox{\vrule width 1.2pt\hskip 3pt
   \vbox{\vskip 6pt}\hskip 3pt\vrule width 0.6pt}\hrule height 0.6pt}\kern1pt}
\magnification 1200

\hoffset=-.1in
\voffset=-.2in
\vsize=7.5in
\hsize=5.6in
\tolerance 10000

\baselineskip 12pt plus 1pt minus 1pt
\pageno=0
\centerline{\bf POINCAR\'E GAUGE THEORIES FOR LINEAL GRAVITY}
\smallskip
%\centerline{\bf Titolo}
\smallskip
\vskip 24pt
\centerline{ G. Grignani}
\vskip 8pt
\centerline{\it Dipartimento di Fisica}
\centerline{\it Universit\'a degli Studi di Perugia}
\centerline{\it I-06100  Perugia -- ITALY}
\vskip 12pt
\centerline{and}
\vskip 12pt
\centerline{ G. Nardelli}
\vskip 8pt
\centerline{\it Dipartimento di Fisica}
\centerline{\it Universit\'a degli Studi di Trento}
\centerline{\it I-38050  Povo (TN) -- ITALY}

\vskip 1.5in
\centerline{Submitted to: {\it Nuclear Physics B}}
\vfill
\vskip -12pt
\noindent  DFUPG-57-1992

\noindent  UTF-266-1992
\hfill August 1992
\eject
\baselineskip 12pt plus 1pt minus 1pt

\centerline{\bf ABSTRACT}
\medskip

We have  shown that two of the most studied models of lineal gravities --
Liouville gravity and a ``string-inspired'' model exhibiting the main
characteristic features of a black-hole solution -- can be formulated as
gauge invariant theories of the Poincar\'e group.
The gauge invariant couplings to matter (particles, scalar and
spinor fields) and  explicit solutions for some matter
field configurations, are provided.
It is shown that both the models, as well as the couplings to matter, can be
obtained as suitable
 dimensional reductions of a $2+1$-dimensional $ISO(2,1)$ gauge invariant
 theory.

\vfill
\eject

\noindent{\bf I.\quad INTRODUCTION}
\medskip
\nobreak

In three and four dimensions, gravity with all its
possible couplings with matter fields and
particles
 can be formulated as a gauge theory of the
Poincar\'e group [1].
However, it is in three dimensions  that the
formulation of gravity as a gauge theory is particularly appealing, as
the Einstein Hilbert action can be written as a pure Chern--Simons term
of the $ISO(2,1)$ gauge potential [2].

In 1+1 dimensions the Einstein tensor $G_{\mu \nu} =R_{\mu \nu} -{1\over
2} g_{\mu \nu} R$ vanishes identically, so that the Einstein field
equations become vacuous.  Some years ago a
class of theories of 2-dimensional gravity, in place of general relativity,
were proposed.
Since then, various models have been studied [3], but two of them seem to
have attracted a particular interest, both for their
simplicity and for their group theoretical properties.
The first one is the so called Liouville gravity [4], and it is the simplest
non trivial theory of lineal gravity based on the scalar curvature $R$. In
this model  $R$ is equated to a cosmological constant $\Lambda$.
To cast this theory into an action principle, an additional
scalar field $\eta$ that acts as a Lagrange multiplier leading to the equation
of motion for $R$, is required. Liouville gravity  then
follows from the action
$$ I_1= \int d^2 x \, \sqrt{-g}\, \eta (R - \Lambda) \ \ . \eqno(1.1)$$
By setting the metric tensor in a conformal form
$g_{\mu \nu} = e^{2\varphi} \eta_{\mu\nu}$, $\eta_{\mu\nu}= {\rm diag}
(1, -1)$ being the
flat Minkowskian metric tensor, the constant curvature condition implies
the Liouville equation $\square \varphi = -{\Lambda \over 2} e^{2\varphi}$
for the conformal field $\varphi$, which justifies the name
of Liouville gravity.

The second model we shall be concerned with has been recently
introduced [5-7] as a
two dimensional toy model leading to an interesting solution that
exhibits all the main characteristic features of a black-hole geometry.
Its action,
$$ I_2 = \int d^2 x \, \sqrt{-g}\, e^{-2\varphi} (R+4g^{\mu \nu}
\partial_\mu
\varphi \partial_\nu \varphi -\Lambda) \ \ , \eqno(1.2)$$
where the scalar field $\varphi$ is usually referred to as the dilaton
field, arises in sigma models deriving from string theory with a two
dimensional  target space and with the antisymmetric tensor field set to
zero. By introducing  the metric tensor $\bar g_{\mu \nu} =e^{2\varphi}
g_{\mu  \nu}$ the action (1.2) can be written in a way that resembles the
action (1.1), namely
$$I_2 = \int d^2 x \, \sqrt{-\bar g}\, (\eta \bar R -\Lambda) \ \ ,
\eqno(1.3)$$
where $\eta =e^{-2\varphi}$ and $\bar R$ is the scalar curvature
corresponding to the metric tensor $\bar g$. The equation of motion
obtained by varying $I_2$ with respect to $\eta$ gives the zero
curvature condition, so that $\bar g_{\mu \nu}$ can be consistently
identified with $\eta_{\mu \nu}$.
The equation for $\eta$  arises by varying $I_2$ with respect to $\bar
g_{\mu \nu}$ and, taking into account the $\bar R=0$ condition, it leads to
$$ \partial_\mu \partial_\nu \eta = -{1\over 2}\Lambda \eta_{\mu \nu}
\ \  .\eqno(1.4)$$
This can be easily integrated
$$ \eta=M-{\Lambda \over 4} (x-x_0)^2   \ \ , \eqno(1.5)$$
$M$ and $x_0$ being integration constants. The {\it
physical} metric
$$ g_{\mu \nu} = {\bar g_{\mu \nu} \over \eta} =
{\eta_{\mu \nu} \over M - {\Lambda\over 4} (x-x_0)^2} \ \ \eqno(1.6)$$
has then the form of a 1+1 dimensional black-hole solution with mass $M$.

Gauge theoretical formulations of the models (1.1) and (1.3)
have been provided
in the recent literature [6,8,9,10]. If we denote by $P_a$ and $J$
the 1+1 dimensional momentum and Lorentz-boost generators
respectively\footnote{$^\dagger$}{In our notation, latin indices
$a,b,c,...=0,1$  denote internal (gauge) indices; they can be raised
and lowered with the Minkowskian metric tensor $\eta_{ab}=\eta^{ab}={\rm
diag}(1, -1).$ The convention on the  antisymmetric symbol
is $\varepsilon^{01} =1$. Greek indices $\mu , \nu
, \rho ,... =0,1$ will always denote space-time  indices.},
Liouville gravity can be presented as a gauge theory [11] of the
de-Sitter group satisfying the algebra
$$ [P_a , J]=\varepsilon_a{}^b P_b \ \ , \qquad \qquad
[P_a , P_b]={\Lambda \over 2}\varepsilon_{ab} J \ \ . \eqno(1.7)$$
The ``black-hole'' model (1.3), instead,  has been recently
formulated as a gauge theory of the  Poincar\'e
group [6] and of a {\it centrally extended} Poincar\'e
group [8,9,10], {\it i.e.} the Poincar\'e group with a central
element $I$ in the momentum algebra
$$ [P_a , J]=\varepsilon_a{}^b P_b \ \ , \qquad \qquad
[P_a , P_b]={i \Lambda \over 2}\varepsilon_{ab} I \ \ . \eqno(1.8)$$

Although these gauge theoretical formulations are consistent and elegant
(the cosmological constant term in Refs.[8-10] naturally arises
from the field strength, and it is not put ``by hand''),
we would like to follow here a different path similar to the one we traced for
 3 and 4 dimensional theories of gravity in Ref.[1].
We want to show in fact that it is possible to describe
1+1 dimensional
 gravity as a gauge theory of the
Poincar\'e group,  as in any other dimension. The Poincar\'e group was also
used in Ref.[6], however, in that formulation, there are some unpleasant
features that were clearly pointed out in Ref.[8]
In Ref.[6] in fact the gauge
transformations have an unconventional form and the  Lagrange density is not
gauge invariant but it changes by a total  derivative. The corresponding
surface term,
 however, cannot be dropped with suitable
asymptotic behaviors of the
gauge fields, as it involves not only
total derivatives of dynamical variables, but also of gauge parameters
that are not required to vanish on the boundary.

Moreover some difficulties arise in trying to couple matter fields to the
centrally extended Poincar\'e theory in a gauge invariant fashion. In fact,
due to the central  extension in the momentum algebra, only infinite
dimensional  representations of the algebra (1.8) are available. On the
contrary as we showed in Ref.[1] a {\it Poincar\'e} gauge theory can be
formulated even in the presence of matter fields, so that it seems more
convenient, when the coupling to matter is under consideration, to use the
Poincar\'e algebra instead of the centrally extended
 one.

To formulate the models as Poincar\'e gauge theories one has to
introduce extra degrees of freedom in the theory: a set of Poincar\'e
coordinates $q^a (x)$ that transform as Poincar\'e vectors under gauge
transformations [1].  These degrees of freedom naturally arise by gauging
the action of a free relativistic particle in Minkowski space so that it
becomes invariant under local Poincar\'e transformations.
However, the presence of the $q^a(x)$ is fundamental in order
to construct a Poincar\'e  gauge theory, not only for describing
gravitational point-particles interaction, but also for matter field
interactions.

The Poincar\'e coordinates $q^a(x)$
 provide a map from the space-time to the
internal (gauge) space ${\cal M}_q$ of coordinates $q^a$ that can be locally
identified with the tangent  space. (See Ref.[1] for an exhaustive
discussion on  such degrees of freedom).

An important point of our formulation is that, contrary to the conventional
formulation of gravity as a gauge theory, the {\it zweibein}  $V^a{}_\mu$  is
not identified with the component $e^a_\mu$ of the gauge potential along the
$P_a$ generators. $V^a{}_\mu$ is given by a  relation involving  the gauge
potential  as well as  the Poincar\'e coordinates $q^a(x)$. This permits
to  avoid the main difficulties encountered in the previous attempts  of
describing these models in the framework of a gauge theory. For  instance we
shall have  Lagrangians that are really gauge invariant, {\it i.e.} that
under gauge transformation do not give rise to total  derivative terms .

Moreover, in the previous gauge theoretical formulations of gravity in $2$
dimensions, the gauge potential is given by
$$A_\mu = V^a{}_\mu P_a + \omega_\mu J \ \ \ , \eqno(1.9)$$
 the {\it zweibein} $V^a{}_\mu$ being the
component of the gauge potential along the translation generators ($P_a$) and
the spin connection $\omega_\mu$ along the Lorentz
generator ($J$).
Consequently, using the Poincar\'e algebra, $V^a{}_\mu$ transforms under
the action of the gauge group as
$$ \delta V^a{}_\mu = -\partial_\mu \rho^a -\varepsilon^a{}_b \omega_\mu
\rho^b + \alpha \varepsilon^a{}_b V^b{}_\mu \ \ , \eqno(1.10)$$
$\alpha$ and $\rho^a$ being the infinitesimal parameter characterizing
the Lorentz boost and translations, respectively.  From Eq. (1.10) it
turns out that the space-time line element
$ds^2 = \eta_{ab} V^a{}_\mu V^b{}_\nu dx^\mu dx^\nu $ is not a scalar
under gauge transformations, {\it i.e.} one has that a ``color singlet''
changes under the action of the gauge group:
$$\delta ds^2 = -\xi^\mu \partial_\mu ds^2 -2[\partial_\mu\xi^\rho
V^a{}_\rho + \xi^\rho {\cal T}^a{}_{\mu \rho}]V_{a \nu}dx^\mu dx^\nu
\ \ , \eqno(1.11)$$
where $\rho^a = V^a{}_\mu \xi^\mu$ and ${\cal T}^a{}_{\mu \nu}
=\partial_\mu V^a{}_\nu -\partial_\nu V^a{}_\mu
+\varepsilon^a{}_b (\omega_\mu V^b{}_\nu -\omega_\nu V^b{}_\mu)$ is the
torsion.
 Nevertheless, one can show
that the gauge variation of $ds^2$ can be compensated by a
diffeomorfism transformation
generated by the vector field $\xi^\mu$ ($\delta x^\mu = \xi^\mu$),
provided the equation of motion of vanishing torsion is taken into
account.\footnote{$^\dagger$}{It has to be noticed, however, that in
principle one might even consider these gauge theoretical models
interacting with matter generating torsion, as it happens for fermions
in higher dimensional theories.}
In our formalism, instead,
 the space-time metric $g_{\mu \nu}= \eta_{ab} V^a{}_\mu V^b{}_\nu $
is always a scalar under gauge transformations, due to the fact that
the {\it zweibein}
$V^a{}_\mu$ is a more complicated combination of gauge potential and
Poincar\'e coordinates transforming as a Lorentz vector under
Poincar\'e gauge transformations.
This feature allows in our approach to couple matter in a gauge
invariant way also in the presence of torsion.

Finally, we would like to mention another property of our
formulation. All the gauge theoretical models describing $1+1$ dimensional
gravity are characterized by the gauge potential equation of motion of
vanishing field strength, that is solved by a pure gauge potential.
A particular type of pure  gauge potential is the vanishing one $A_\mu=0$,
which leads, in the conventional approach to gravity as a gauge theory,
to the degenerate solution $V^a{}_\mu=0$ (see Eq. (1.9)).
Therefore, in this case the
usual geometrical interpretation of $V^a{}_\mu$ is lost. On the
contrary, following our approach, the gauge choice $A_\mu =0$ is
perfectly acceptable. It corresponds to  absence of interaction, namely to
free matter actions, as it happens in Yang--Mills theory.

In Sect.IV we also show that both the Poincar\'e gauge
theories describing the lineal models (1.1) and (1.3) can be obtained by
 suitable dimensional reductions of the  $ISO(2,1)$ gauge
theory in  $2+1$ dimensions (see also Ref. [10]).

In Sect.II we formulate the
models in an $ISO(1,1)$ gauge invariant way and we
 discuss the connection with the usual
Einsteinian formulation  Eqs. (1.1) and (1.3).
In Sect.III we provide the $ISO(1,1) $
gauge invariant  couplings with matter (particles, scalar and
spinor fields). The equations of motion will be explicitly solved
for some  matter fields configurations. We provide in fact new solutions
of the $ISO(1,1)$ gauge theory coupled to
 fermion, scalar and point particle sources.

In Sect.IV we also discuss the dimensional reductions that lead from a
general $2+1$-dimensional $ISO(2,1)$ gauge invariant theory with matter,
to the $ISO(1,1)$ gauge invariant actions for gravity and matter in
$2$-dimensions. An $ISO(2,1)$ dimensional reduction leading to the
de-Sitter-Liouville  theory described in Refs. [5] is also discussed.

In Sect.V we draw our conclusions,
whereas in the Appendix some technical points concerning the value of
the {\it zweibein} in the pure gauge solutions are discussed.

\bigskip

\noindent{\bf II. \quad ISO(1,1) GAUGE INVARIANT MODELS }
\medskip
\nobreak

We shall formulate our models by gauging the corresponding $ISO(1,1)$ global
invariant actions. In this way, the relations between the gauge
potential and the relevant physical quantities of the theory will
naturally emerge.
Let us consider the action of
a free relativistic particle in a two
dimensional Minkowskian manifold ${\cal M}_q$,
$$ S_{\rm free} = \int d\tau \, [p_a \dot q^a + \lambda (p^2 - m^2)] \ \
, \eqno(2.1)$$
where $(q^a, p_b)$ are canonically conjugate variables, $\tau$ denotes
the proper time of the particle and $\lambda$ is a Lagrange multiplier
introduced to enforce the constraint $p^2 =m^2$, $m$ being the mass of
the particle.  $S_{\rm free}$ is obviously invariant under global
Poincar\'e transformations
$$\eqalignno{\delta q^a &=\alpha \varepsilon^a{}_bq^b +\rho^a \ \ ,
&\hbox{(2.2a)}\cr
\delta p_a & = \alpha \varepsilon_a{}^b p_b \ \ , &\hbox{(2.2b)} \cr }
$$
where $\alpha$ and $\rho^a$ are constant infinitesimal parameters
associated to the Poincar\'e generators $J$ and $P_a$ satisfying the
algebra
$$ [P_a, P_b]=0 \ , \qquad \qquad [P_a, J]=\varepsilon_a{}^b P_b \ \ .
\eqno(2.3)$$
As long as the particle is free, the canonical variables ($q^a, p_b)$ can be
identified with the space-time canonical variables $(x^\mu , \pi_\nu)$,
because in this case the space-time is Minkowskian.
If instead we include gravitational interactions such identification does
no longer hold and the  $q^a$ and $p_b$ variables should be considered as
functions of the space-time  trajectory of the particle $x^\mu (\tau)$.

To make the $ISO(1,1)$ symmetry local,
 the proper time derivative
$\dot q^a$ must be replaced by a  covariant derivative
${\cal D}_\tau q^a = {\cal D}_\mu q^a (x) \dot x^\mu$
transforming as $p_a$, where it is assumed
that the $q^a (x)$ depend on the proper time $\tau$ only through the
space-time trajectory $x^\mu (\tau$).
The covariant derivative must contain an homogeneous part, with gauge
potential $\omega_\mu$ associated to the Lorentz transformation $J$,
and an inhomogeneous part with gauge potential $e^a{}_\mu$ associated to
the translation generators $P_a$, whose presence is necessary
in order to reabsorb the
inhomogeneous term in the gauge transformation (2.2b).
Thus we look for a covariant derivative of the form
$${\cal D}_\mu q^a = \partial_\mu q^a +\varepsilon^a{}_b \omega_\mu q^b
+ e^a{}_\mu \ \ . \eqno(2.4)$$
By imposing the condition $\delta {\cal D}_\mu q^a = \alpha
\varepsilon^a{}_b {\cal D}_\mu q^b$,
 we get the transformation
laws for the gauge potentials
$$\eqalignno{\delta \omega_\mu & =-\partial_\mu \alpha \ \ ,
&\hbox{(2.5a)} \cr \delta e^a{}_\mu & = -\partial_\mu \rho^a
-\varepsilon^a{}_b \omega_\mu \rho^b +\alpha \varepsilon^a{}_b e^b{}_\mu
\ \ . &\hbox{(2.5b)} \cr}$$
By construction, the  $ISO(1,1)$ gauge invariant action for a particle
then reads
$$S_{\rm part} = \int d\tau \, [p_a {\cal D}_\mu q^a \dot x^\mu +
\lambda (p^2 - m^2)] \ \ , \eqno(2.6)$$
where $\dot x^\mu (\tau)$ is the vector tangent to the particle
trajectory, and $q^a (x(\tau))$ describes an image trajectory in the
internal Poincar\'e space.

By comparing Eq.(2.6)  with the usual action for a particle in the
space-time written in the first order formalism , it turns out that the
physical {\it zweibein}  is given by
$$ V^a{}_\mu = {\cal D}_\mu q^a =
\partial_\mu q^a +\varepsilon^a{}_b \omega_\mu q^b
+ e^a{}_\mu \ \ . \eqno(2.7)$$

Usually, in the framework of a Poincar\'e gauge
theory, the component $e^a{}_\mu$ of the gauge potential along the $P_a$
generator is interpreted as the physical {\it zweibein}.
As Eq.(2.7) shows, in our approach this statement is incorrect. However,
since a choice of the map $q^a(x)$ corresponds to a gauge choice that fixes the
translational part of the Poincar\'e symmetry, leaving the theory
invariant under residual Lorentz transformations, there exists a gauge
choice in which the {\it zweibein} can be indeed identified with
$e^a{}_\mu$. In fact, if we choose the
``physical'' gauge condition $q^a =0$ we have
from Eq.(2.7) $V^a{}_\mu = e^a{}_\mu$.
Hence, this interpretation for the gauge potential $e^a{}_\mu$
 only holds in a
particular gauge choice
with a residual $SO(1,1)$ gauge freedom
 and, consequently, in the framework of a {\it
Lorentz} gauge theory.
Alternatively, if we keep the whole Poincar\'e symmetry or if we
choose a gauge different from the physical one, the {\it zweibein} has
the more complicated structure given in Eq.(2.7).

This feature allows to maintain a close analogy with any ordinary
Yang--Mills-type  gauge theory. Just as an example, suppose we have
field equations of vanishing field strength, so that the gauge potential
is a pure gauge. Then one can choose as a particular pure gauge solution
the vanishing connection $e^a{}_\mu = \omega_\mu =0$ that fixes
completely the gauge arbitrariness. Within this gauge choice the {\it
zweibein} is given everywhere by $\partial_\mu q^a$ which, being the
$q^a$ Minkowskian coordinates, simply implies that the space-time is
flat and the $q^a$ can be globally interpreted as space-time
coordinates. [See the Appendix for a more exhaustive discussion on this
point].

Next, we turn to the gauge potential Lagrangian.
The transformations (2.5) are the usual gauge
transformations one expects from a standard non Abelian gauge theory:
introducing the Lie algebra valued gauge potential $A_\mu$ and
the infinitesimal parameter $u$
$$ \eqalignno{A_\mu &= e^a{}_\mu P_a + \omega_\mu J \ \ , &\hbox{(2.8a)}
\cr u &= \rho^a P_a + \alpha J \ \ , &\hbox{(2.8b)} \cr}$$
the transformations (2.5) acquire the familiar form
$$ \delta A_\mu = -\partial_\mu u - [A_\mu  , u] \equiv - \Delta_\mu u \ \ ,
\eqno(2.9)$$
and the Lie algebra valued field strength
$$ \eqalign{F_{\mu \nu} &= [\Delta_\mu , \Delta_\nu] = P_a T^a{}_{\mu
\nu} + J R_{\mu \nu} \cr
& =P_a [\partial_\mu e^a{}_\nu - \partial_\nu e^a{}_\mu + \varepsilon^a{}_b
(\omega_\mu e^b{}_\nu - \omega_\nu e^b{}_\mu)] + J[\partial_\mu
\omega_\nu - \partial_\nu \omega_\mu] \cr} \eqno(2.10)$$
transforms covariantly
 under gauge transformations, namely
$$\delta F_{\mu \nu} =  -[F_{\mu \nu} , u]
= P_a \varepsilon^a{}_b (\alpha T^b{}_{\mu \nu} - \rho^b R_{\mu \nu})
\eqno(2.11)$$
The finite version of Eq.(2.11) can be conveniently written by
introducing the triplet $F^A{}_{\mu \nu} \equiv (T^a{}_{\mu \nu}, R_{\mu
\nu})$, $A=(a,2)$; then, integrating
Eq.(2.11), $F^A{}_{\mu \nu}$ transforms according
to the three dimensional adjoint representation of the Poincar\'e group,
$$\eqalign{F^A{}_{\mu \nu} \longrightarrow &
(U^{-1})^A{}_B F^B{}_{\mu \nu} \cr \noalign{\vskip .2cm}
U^A{}_B =& \left(\matrix{M^a{}_b&\varepsilon^a{}_c\rho^c\cr
0&1\cr}\right) \cr} \eqno(2.12)$$
where $M^a{}_b$ is the finite boost
$$ M^a{}_b = \delta^a{}_b {\rm cosh}\alpha - \varepsilon^a{}_b {\rm
sinh} \alpha \ \ . \eqno(2.13)$$
Consequently, the Lagrangian density
$${\cal L}= \varepsilon^{\mu \nu} \eta_A F^A{}_{\mu \nu} \ \
\eqno(2.14)$$
is gauge invariant provided the Lagrange multiplier triplet $\eta_A=(\eta_a,
\eta_2) $
transforms according to the coadjoint representation
$$ \eta_A \longrightarrow \eta_B U^B{}_A \ \  , \eqno(2.15)$$
or, in infinitesimal form,
$$ \eqalign{\delta \eta_a &= \alpha \varepsilon_a{}^b \eta_b \cr
\delta \eta_2 & =\varepsilon^a{}_b \rho^b \eta_a \cr} \ \ . \eqno(2.16)$$

It should be remarked that, since $e^a{}_\mu$ is not the physical {\it
zweibein}, $T^a{}_{\mu \nu}$ cannot be interpreted as the physical
torsion (that we shall denote  by ${\cal T}^a{}_{\mu \nu}$); as a matter of
fact, from Eq.(2.7) one can easily find that ${\cal T}^a{}_{\mu \nu}$
can  be written
as $$ {\cal T}^a{}_{\mu \nu} = T^a{}_{\mu \nu} + \varepsilon^a{}_b q^b
R_{\mu \nu} \ \ , \eqno(2.17)$$
and, as expected, only in the physical gauge
${\cal T}^a{}_{\mu \nu} \equiv T^a{}_{\mu \nu}$. However,
 since we are interested in a formulation of lineal gravity as a gauge
theory of the Poincar\'e (and not Lorentz) group, we shall not choose
any specific value for the Poincar\'e variables $q^a$.

By means of the covariant derivatives ${\cal D}_\mu q^a$, one can
also construct an $ISO(1,1)$ gauge invariant cosmological constant term, namely
$$ \Lambda \varepsilon^{\mu \nu} \varepsilon_{ab} {\cal D}_\mu q^a
{\cal D}_\nu q^b \ \ , \eqno(2.18)$$
$\Lambda $ being an arbitrary constant.

We shall now show that the ``black-hole'' model
Eq. (1.3) can be formulated  in terms of the following
 $ISO(1,1)$ gauge invariant action
$$ S_{\rm BH} = \int d^2x \, \varepsilon^{\mu \nu} \left[
\eta_A F^A{}_{\mu \nu} + {\Lambda \over 2} \varepsilon_{ab} {\cal D}_\mu
q^a {\cal D}_\nu q^b \right] \ \ , \eqno(2.19)$$
with $\eta$ replaced by $\eta_2 -\varepsilon_{ab} \eta^a q^b$.

In fact, by varying $S_{\rm BH}$ with respect to $\eta_A$ we get the zero
field strength condition
$$ T^a{}_{\mu \nu} = R_{\mu \nu} =0 \ \ , \eqno(2.20{\rm a})$$
whereas the equations of motion obtained by varying the action with
respect to $\omega_\nu$, $e^a{}_\nu$ and $q^a$ are, respectively,
$$\eqalignno{\partial_\mu \eta_2 + \varepsilon^a{}_b \eta_a e^b{}_\mu
-{\Lambda\over 2} q_a {\cal D}_\mu q^a &=0 \ \ , &\hbox{(2.20b)} \cr
\noalign{\vskip .2cm} \partial_\mu\eta_a +\omega_\mu \varepsilon_a{}^b
\eta_b + {\Lambda\over 2}\varepsilon_{ab}{\cal D}_\mu q^b &=0 \ \ ,
&\hbox{(2.20c)} \cr \noalign{\vskip .2cm}
{\cal T}^a{}_{\mu \nu} = T^a{}_{\mu \nu} + \varepsilon^a{}_b q^b R_{\mu
\nu} &=0 \ \ . &\hbox{(2.20d)} \cr} $$

In solving these equations, we first notice  that Eq.(2.20d) of vanishing
``physical'' torsion is automatically satisfied by the  zero field
strength condition Eq.(2.20a). This is a general feature of the
Poincar\'e gauge theories describing gravity in any dimensions: in
order to formulate gravity as an $ISO(n,1)$
gauge theory in an $n+1$ dimensional space-time, one has to
introduce the extra degrees of freedom (Poincar\'e coordinates) $q^a$
and, consequently, one has extra equations of motion. However, these
extra equations are harmless as they  either
provide identically satisfied conditions (as in the case of Eq.(2.20d))
or collapse with other equations (see Ref.[1]).

Next, we solve the remaining equations, and to this purpose we have to
choose a gauge.
 The physical gauge $q^a =0$  is very useful to
establish contact with the  models previously analyzed in the
literature; in fact it can be easily shown that in the physical gauge
the field equations (2.20) essentially collapse to those analyzed in Ref.[6].
However, there is a gauge choice in which the equations are even
simpler: since the field strength vanishes, we can choose
the zero connection condition
$e^a{}_\mu = \omega_\mu =0$, which is consistent, up to a constant,
 with the choice $q^a=\delta^a{}_\mu x^\mu$.
Consequently the {\it zweibein} is a Kronecker delta, the metric
tensor is Minkowskian and Eqs. (2.20) become
$$\eqalign{\partial_\mu \eta_2 & ={\Lambda \over 2} x_\mu \cr
\noalign{\vskip .2cm} \partial_\mu \eta_a & ={\Lambda \over 2}
\varepsilon_{\mu a} \cr } \eqno(2.21)$$
It is straightforward to see that
the general solution of  the equation for
$\eta= \eta_2 -\varepsilon_{ab}\eta^a q^b$
is then given by Eq.(1.5) (up to a constant translation,
which is a consequence of the arbitrariness $q^a \to q^a + {\rm const.}$),
so that the action (2.19) reproduces the two dimensional
black-hole
 metric, once the relation between  the $\eta$ field and the metric
tensor Eq.(1.6) is taken into account.
Had we chosen the physical gauge $q^a=0$, the equations for $\eta_a$ and
$\eta_2$ would have been coupled, leading therefore to a second order
equation for $\eta_2$, namely Eq.(1.4), and in this case $\eta_2$ would
coincide with $\eta$.

Let us turn now to the $ISO(1,1)$ gauge formulation of Liouville gravity. A
Poincar\'e gauge invariant action that leads to the equation of motion of
Liouville gravity is
$$ \tilde S= \int d^2x \, \varepsilon^{\mu \nu} \bigl[
\eta_A F^A{}_{\mu \nu}  +{\Lambda\over 2} (\eta_2
-\varepsilon^c{}_d\eta_c q^d)
 \varepsilon_{ab} {\cal D}_\mu q^a {\cal D}_\nu q^b
\bigr] \ \ . \eqno(2.22)$$

In fact, the field equations obtained by varying $\tilde S$ with respect
to $\eta_2$, $\eta_a$, $\omega_\nu$, $e^a{}_\nu$ and $q^a$ are given
by, respectively
$$\eqalignno{\varepsilon^{\mu \nu} R_{\mu \nu} + {\Lambda\over 2}
\varepsilon^{\mu \nu} \varepsilon_{ab} {\cal D}_\mu q^a {\cal D}_\nu q^b
&=0 \ \ , &\hbox{(2.23a)} \cr
\varepsilon^{\mu \nu} T^a{}_{\mu \nu} - {\Lambda\over 2} \varepsilon^a{}_b q^b
\varepsilon^{\mu \nu} \varepsilon_{cd} {\cal D}_\mu q^c {\cal D}_\nu q^d
&=0 \ \ , &\hbox{(2.23b)} \cr
\partial_\mu \eta_2 + \varepsilon^a{}_b \eta_a e^b{}_\mu -
{\Lambda\over
2} (\eta_2 - \varepsilon^a{}_b \eta_a q^b) q_c{\cal D}_\mu q^c &=0 \ \ ,
&\hbox{(2.23c)} \cr
\partial_\mu \eta_a+\varepsilon_a{}^b \omega_\mu \eta_b +
{\Lambda\over 2}\varepsilon_{ab} {\cal D}_\mu q^b (\eta_2 -\varepsilon^c{}_d
\eta_c q^d) &=0 \ \ , &\hbox{(2.23d)} \cr
\varepsilon^{\mu \nu} \varepsilon_{ab} \left[
2{\cal D}_\mu q^b \partial_\nu (\eta_2 -\varepsilon^c{}_d \eta_c q^d)
+\eta^b \varepsilon_{cd} {\cal D}_\mu q^c{\cal D}_\nu q^d  \right]&=0 \
\ , &\hbox{(2.23e)} \cr}$$
where in the last equation we took  the vanishing physical
torsion condition into account. If we add Eq.(2.23c) to
Eq.(2.23d) multiplied by $\varepsilon_{ca}q^c$ we get Eq.(2.23e): as
expected, the field equation corresponding to the variation of
$\tilde S$ with respect to $q^a$  provides an identically satisfied
condition. That these equations reproduces Liouville gravity can be
 easily realized by noticing that the scalar curvature $R$ expressed in terms
of the gauge potentials and Poincar\'e coordinates reads
$R=\varepsilon^{\mu \nu} R_{\mu \nu}  / {\rm det} ({\cal D}_\rho q^c)$,
so that Eq.(2.23a) can be rewritten, assuming the invertibility of the
{\it zweibein}, as the familiar constant curvature condition
$$R=\Lambda \ \ \eqno(2.24)$$ leading to Liouville gravity.

Of course Eq.(2.22) is not the only possible gauge invariant action that
can be constructed, expecially in view of the fact that the quantity
$\eta_2 - \varepsilon^a{}_b \eta_a q^b$ is dimensionless and gauge invariant.
Nevertheless Eq.(2.22) provides the $ISO(1,1)$ gauge invariant model leading
to Liouville gravity.

 Explicit solutions can be found by choosing a
particular gauge. In this  case we shall choose the physical gauge $q^a =0$,
so that the {\it  zweibein} can be identified with the $e^a{}_\mu$ component
of the gauge potential. By choosing conformal coordinates $e^a{}_\mu =
\delta^a{}_\mu e^\varphi$ Eqs.(2.23)  rearrange as $$\eqalignno{2\square
\varphi &= -\Lambda e^{2\varphi} \ \ ,  &\hbox{(2.25a)} \cr
\omega_\mu&=\varepsilon^\nu{}_\mu \partial_\nu \varphi \ \  ,&\hbox{(2.25b)}
\cr \partial_\mu \eta_2 &= \varepsilon_\mu{}^\nu \eta_\nu e^\varphi\ \ ,
&\hbox{(2.25c)} \cr \partial_\mu \eta_\nu +\varepsilon_\nu{}^\rho
\varepsilon^\sigma{}_\mu \partial_\sigma \varphi \eta_\rho & = {\Lambda
\over 2}\varepsilon_{\mu \nu}  \eta_2 e^\varphi \ \ , &\hbox{(2.25d)} \cr} $$
where $\square$ is the Minkowskian d'Alembertian and
we omitted Eq.(2.23e) as redundant. Given a solution of the
Liouville equation (2.25a), the value of the  connection $\omega_\mu$ and
 lagrange multiplier $\eta_a$ and $\eta_2$ has
to be chosen consistently with Eqs.(2.25b - d).
The general solution of Eq.(2.25a) [9] would involve two arbitrary
functions; however, the residual coordinate invariance
 that the conformal choice
entails, allows to choose
$$ e^{2\varphi} =
{1\over \left(1+{\Lambda\over 8} x^2 \right)^2}  \ \ ,
\eqno(2.26)$$
which lead to the following value for the Lagrange multiplier  $\eta_2$
$$ \eta_2 = {\alpha_a x^a + \alpha_2 \left(1 - {\Lambda \over 8} x^2
\right) \over 1+{\Lambda\over 8}x^2 } \ \ , \eqno(2.27)$$
$\alpha_a$ and $\alpha_2$ being arbitrary constants.
 The corresponding values for $\omega_\mu$ and $\eta_a$
 can be straightforwardly
derived from Eq.(2.25b,c).
Eq.(2.26) gives the metric of the space-time in Liouville gravity,
$$g_{\mu \nu}={\eta_{\mu \nu}
\over \left(1+{\Lambda\over 8} x^2 \right)^2}  \ \ .\eqno(2.28) $$

\bigskip

\noindent{\bf III. \quad ISO(1,1) GAUGE INVARIANT COUPLINGS TO MATTER }
\medskip
\nobreak

In this Section we shall provide all the relevant gauge invariant
couplings to matter (particles, scalar and spinor fields).
It is important to remark that the formulation of black-hole gravity as
a gauge theory has been performed {\it after} the action (1.2) was
reduced to the simplified form (1.3), by redefining  the metric through
a conformal rescaling. Such a rescaling in the presence of matter gives
rise to undesirable non-linear couplings  of the dilaton field $\varphi$
to matter fields (or particles), unless the matter under consideration
is conformally invariant. Therefore, the gauge invariant quantity
$\eta=\eta_2 -\varepsilon_{ab} \eta^a q^b$ can be interpreted as the
conformal factor of a $2$-dimensional black-hole model interacting with
matter only
when massless scalar and spinor fields are considered.
 Nevertheless, it is instructive to see how a
Poincar\'e gauge invariant coupling to matter can be achieved in a more
general context, and for this reason we shall also include in this Section
 the interactions with particles and massive (scalar and spinor)
fields. Such non-conformally invariant matter configurations will be no
longer related to the black-hole model but, rather, they should be
regarded  more generally as Poincar\'e gauge invariant couplings to the
action (2.19).
In the case of fields, however, the massless limit of the solutions we
shall provide is well defined and the corresponding metrics can indeed
be interpreted as $2$ dimensional black holes interacting with conformal
matter.

We begin with the coupling to particles. We then consider the
$ISO(1,1)$ gauge invariant action
$$ S = \int d^2x \, \varepsilon^{\mu \nu} \left[
\eta_A F^A{}_{\mu \nu} + {\Lambda \over 2} \varepsilon_{ab} {\cal D}_\mu
q^a {\cal D}_\nu q^b \right] +
\alpha\int d\tau \, [p_a {\cal D}_\mu q^a \dot x^\mu +
\lambda (p^2 - m^2)] \ \ , \eqno(3.1)$$
$\alpha$ being a dimensionless constant.
The equations of motion deriving from $S$ are
$$\eqalignno{ {\delta S \over \delta \eta_A} =0 \quad & \rightarrow
\quad  F^A{}_{\mu \nu} =0 \ , &\hbox{(3.2a)} \cr
 {\delta S \over \delta \omega_\nu} =0 \quad &\rightarrow
\quad  \partial_\mu \eta_2 + \varepsilon^a{}_b \eta_a e^b{}_\mu
-{\Lambda \over 2} q_a {\cal D}_\mu q^a &\cr
& \qquad \quad
 +{\alpha \over 2} \varepsilon_{\mu \nu} \varepsilon^a{}_b \int d\tau
p_a q^b \dot x^\nu \delta^{(2)} (x -x(\tau)) =0 \  , &\hbox{(3.2b)} \cr
{\delta S \over \delta e^a{}_\nu} =0 \quad &\rightarrow
\quad \partial_\mu \eta_a +\varepsilon_a{}^b \eta_b \omega_\mu
+{\Lambda\over 2} \varepsilon_{ab}{\cal D}_\mu q^b &\cr&\qquad \quad
 +{\alpha \over 2} \varepsilon_{\mu \nu}  \int d\tau
p_a  \dot x^\nu \delta^{(2)} (x -x(\tau)) =0 \  , &\hbox{(3.2c)} \cr
{\delta S \over \delta q^a} =0 \quad &\rightarrow
\quad \Lambda \varepsilon^{\mu \nu} \varepsilon_{ab} {\cal T}^b{}_{\mu
\nu} +2\alpha \int d\tau (\partial_\mu p_a +\varepsilon_a{}^b \omega_\mu
p_b) \dot x^\mu \delta^{(2)} (x -x(\tau)) =0 \ , &\hbox{(3.2d)} \cr
{\delta S \over \delta p_a} =0 \quad &\rightarrow
\quad {\cal D}_\mu q^a \dot x^\mu (\tau) + 2\lambda p^a (\tau) =0
 \ , &\hbox{(3.2e)}\cr
{\delta S \over \delta \lambda} =0 \quad &\rightarrow
\quad  p^2 - m^2 =0 \ , &\hbox{(3.2f)} \cr
{\delta S \over \delta x^\mu (\tau)} =0 \quad &\rightarrow
\quad p_a \dot x^\nu {\cal T}^a{}_{\mu \nu} =0 \ . &\hbox{(3.2g)} \cr
}$$
As expected, one of the equations (the last one) is automatically
satisfied. Eq.(3.2a) implies the zero curvature and torsion condition,
so that we can consistently choose the vanishing connection gauge
$\omega_\mu = e^a{}_\mu =0$ which, in turn, implies
that the flat space ${\cal M}_q$ with coordinates $q^a$
coincide with the
space-time with coordinates $x^\mu$.
The Eqs.(3.2d-f) are then equivalent to the equations of motion of a free
relativistic particle in a Minkowskian space-time, leading to the
solution
$$\eqalignno{p^a &=(m\gamma , m\gamma v) \ \ , &\hbox{(3.3a)}\cr
 x^a (x^0)& = (x^0, vx^0) \ \ , &\hbox{(3.3b)}\cr
\lambda &=-{1\over 2m\gamma} \ \ , &\hbox{(3.3c)}\cr}$$
where $v$ is the velocity of the particle,
$\gamma = (1-v^2)^{-1/2}$, $ x^a (x^0)$ denotes the particle
trajectory and we chose $x^0=\tau$ and
 $x^a ( \tau =0) \equiv 0$. By substituting these solutions in the field
equations   for the Lagrange multipliers $\eta_A$
we arrive at
$$ \eqalign{\partial_\mu \eta_2 & ={\Lambda \over 2} x_\mu \cr
\noalign{\vskip .2cm} \partial_\mu \eta^a & ={\Lambda \over 2}
\varepsilon_{\mu}{}^\nu \delta^a{}_\nu -
{\alpha \over 2} \varepsilon_{\mu \nu} p^a
{d\bar x^\nu \over dx^0} \delta(x^1 -v x^0)
 \cr }, \eqno(3.4)$$
namely the equation for $\eta_2$ is
unchanged if compared with
the previous case without particle interaction
 (see Eq.(2.21)) and consequently its solution
is again given by Eq.(1.5). The solution for $\eta^a$ is instead given
by
$$\eta^a = {\Lambda \over 2}\varepsilon_\mu{}^\nu  \delta^a{}_\nu x^\mu
 -{\alpha \over 2} p^a \vartheta
(x^1 -vx^0) + C^a \ \ , \eqno(3.5)$$
where $C^a$ is an arbitrary Minkowskian
 constant vector and $\vartheta$ denotes the
Heavyside function.

Next we consider couplings with matter fields (scalar and spinor). We
 require that the matter fields carry a representation of the
Poincar\'e group. Let us consider first the scalar field. In order to
formulate a metric independent theory we need to work in a first order
formalism. We then introduce a  field $\varphi^A =(\varphi^a, m^2 \varphi) $,
[ $A=0,1,2$ and  $m$ being a mass parameter]
 carrying a vectorial ($3 \times 3$) representation of the
Poincar\'e generators, {\it i.e.} transforming as
$$ \delta \varphi^A = [\alpha J + \rho^a P_a]^A{}_B \varphi^B \ \ ,
\eqno(3.6) $$
where the vectorial representation of the Poincar\'e algebra is given by
$$ J^A{}_B = \left(\matrix{\varepsilon^a{}_b & 0\cr 0&0} \right)
\ , \qquad (P_a)^A{}_B = \left( \matrix{0&\delta^a{}_b\cr 0&0} \right)
\ \ . \eqno(3.7)$$
Accordingly we can construct the covariant derivative
$$ {\cal D}_\mu \varphi^A = (\partial_\mu {\bf 1} + \omega_\mu J + e^a{}_\mu
P_a )^A{}_B \varphi^B \eqno(3.8)$$
or, in components
$$ \eqalign{{\cal D}_\mu \varphi^a &= \partial_\mu\varphi^a+\varepsilon^a{}_b
\omega_\mu \varphi^b + m^2 e^a{}_\mu \varphi \cr
{\cal D}_\mu \varphi & = \partial_\mu \varphi \cr} \eqno(3.9)$$

An $ISO(1,1)$ gauge invariant action is given by
$$ S_S=\int d^2x \, \varepsilon^{\mu \nu} \varepsilon_{ab}{\cal D}_\mu q^a
\left[ {\cal D}_\nu \varphi^b \varphi - {\cal D}_\nu \varphi \varphi^b
+ {\cal D}_\nu q^b (\varphi_c -m^2 \varphi q_c) (\varphi^c -m^2 \varphi
q^c) \right] \ \ . \eqno(3.10)$$
The variation of $S_S$ with respect to $\varphi^a$  gives a relation
between $\varphi$ and $\varphi^a$ that substituted in the equation of
motion for $\varphi$ and assuming the invertibility of the {\it
zweibein} leads to the Klein-Gordon equation in curved space-time
for a scalar field $\varphi$ with mass $2m$.
Hence, if we want to couple
in a gauge invariant fashion a scalar field to the  action
(2.19) we are led to
$$\eqalign{ S &= \int d^2x \, \varepsilon^{\mu \nu} \left[
\eta_A F^A{}_{\mu \nu} + \left({\Lambda \over 2}
+\alpha (\varphi_c -m^2 \varphi q_c)(\varphi^c -m^2 \varphi q^c)
\right)\varepsilon_{ab} {\cal D}_\mu
q^a {\cal D}_\nu q^b \right. \cr  &\left.
+ \alpha \varepsilon_{ab}{\cal D}_\mu q^a
\left( {\cal D}_\nu \varphi^b \varphi - {\cal D}_\nu \varphi
\varphi^b\right){ \over }
\right] \cr} \eqno(3.11)$$
$\alpha$ being a dimensionless constant. The equations of motion
deriving from the action (3.11) are
$$\eqalignno{ {\delta S \over \delta \eta_A} =0 \quad  \rightarrow
\quad & F^A{}_{\mu \nu} =0 \ , &\hbox{(3.12a)} \cr
 {\delta S \over \delta \omega_\nu} =0 \quad \rightarrow
\quad & \partial_\mu \eta_2 + \varepsilon^a{}_b \eta_a e^b{}_\mu
-\left({\Lambda \over 2} +\alpha (\varphi_b -m^2 \varphi q_b)
(\varphi^b -m^2 \varphi q^b)\right)
 q_a {\cal D}_\mu q^a &\cr & -{\alpha\over 2}\varphi \varphi_a {\cal D}_\mu q^a
 -{\alpha \over 2}q_a[\varphi {\cal D}_\mu \varphi^a - {\cal D}_\mu
\varphi \varphi^a]=0 \  , &\hbox{(3.12b)} \cr
{\delta S \over \delta e^a{}_\nu} =0 \quad \rightarrow
\quad & \partial_\mu \eta_a +\varepsilon_a{}^b \eta_b \omega_\mu
+\left({\Lambda\over 2} +\alpha (\varphi_c -m^2 \varphi q_c)
(\varphi^c -m^2 \varphi q^c)\right)\varepsilon_{ab}{\cal D}_\mu q^b
&\cr & + m^2 {\alpha \over 2}
\varepsilon_{ab} {\cal D}_\mu q^b \varphi^2
+{\alpha \over 2} \varepsilon_{ab} [\varphi {\cal D}_\mu \varphi^b -{\cal
D}_\mu \varphi \varphi^b] =0
 \  , &\hbox{(3.12c)} \cr
{\delta S \over \delta \varphi^a} =0 \quad \rightarrow
\quad & \varepsilon^{\mu \nu} {\cal D}_\mu q^b [\varepsilon_{bc} {\cal
D}_\nu q^c (\varphi_a - m^2 \varphi q_a) + \varepsilon_{ab} {\cal D}_\nu
\varphi ]=0
 \ , &\hbox{(3.12d)} \cr
{\delta S \over \delta \varphi} =0 \quad \rightarrow
\quad & {\cal D}_\nu \varphi^b -m^2 q_c (\varphi^c -m^2 \varphi q^c)
{\cal D}_\nu q^b=0
 \ , &\hbox{(3.12e)}\cr }$$
where we omitted the equation obtained by varying $S$ with respect to
$q^a$ as redundant.
To solve Eqs. (3.12) we shall choose the gauge $q^a =0$, $\omega_\mu=0$,
$e^a{}_\mu=\delta^a{}_\mu$, which is consistent with Eq. (3.12a).
Then, decoupling  Eqs.( 3.12d,e) we get the Klein-Gordon equation
for the scalar field $\varphi$,
$$ \square \varphi + 4m^2 \varphi =0 \ \ , \eqno(3.13)$$
$\square$ being the Minkowskian d'Alembertian. Eq. (3.13)
 is obviously solved by
plane waves
$$ \varphi^{(\pm)} = e^{\mp i kx} \ , \qquad k^2 = 4m^2\ , \qquad
k_0 = \sqrt{ (k_1)^2 + (2m)^2} >0 \ \ . \eqno(3.14)$$
We shall choose as an  example the simplest superposition of the
solutions (3.14)  consistent with the reality properties of the
scalar field $\varphi$, namely $\varphi= \cos kx$.

The value of $\varphi^a$ has to be chosen according to Eq.(3.12d),
 {\it i.e.} $\varphi_a  ={1\over 2}\delta_a{}^\mu
\partial_\mu \varphi = -{1\over 2} k_a \sin kx$.

Finally, the equations for the Lagrange multipliers in our gauge choice
become
$$\eqalign{\partial_\mu \eta_2 &= \varepsilon_{\mu \nu} \delta^\nu{}_a
\eta^a + {\alpha \over 2}  \varphi \varphi_a \delta^a{}_\mu \cr
&=\varepsilon_{\mu \nu} \delta^\nu{}_a
\eta^a  - {\alpha \over 8} k_\mu \sin 2kx \cr
\partial_\mu \eta_a &= \varepsilon_{ba} \left[ {\alpha \over 2} (\varphi
\partial_\mu \varphi^b -\partial_\mu \varphi \varphi^b) +\delta^b{}_\mu
\left( {\Lambda \over 2} +\alpha (\varphi_c \varphi^c +m^2 \varphi^2)
\right) \right] \cr
&= \varepsilon_{ba} \left[ \delta^b{}_\mu \left({\Lambda \over 2}
+\alpha m^2 \right) -{\alpha\over 4} \delta^b{}_\nu k_\mu k^\nu \cos 2kx
\right] \cr} \eqno(3.15)$$
whose solutions are given by
$$\eqalign{ \eta^a &= {\delta^a{}_\nu \varepsilon_\mu{}^{\nu} \over 2}
\left[\Lambda x^\mu - {\alpha \over 2} \bigl( (kx) k^\mu - k^2 x^\mu
\bigr) \right] \cr
\eta_2 & = -{\Lambda \over 4} x^2 + {\alpha \over 8}\left[
(kx)^2 - k^2 x^2 + {1\over 2} \cos 2kx \right] + D \cr}\eqno(3.16)$$
$D$ being an integration constant.

To include fermions, we have to introduce $2\times 2$ Dirac matrices
$\gamma^a$ normalized as $\{ \gamma^a , \gamma^b \} = 2 \eta^{ab}$.
We shall choose $\gamma^a = (\sigma^3 , i\sigma^2)$,
and $\gamma^5 = \gamma^0 \gamma^1 = \sigma^1$, $\sigma^i$ being the
Pauli matrices.
Then, a spinorial representation of the Poincar\'e algebra with a
non-vanishing
translation generator is $$ J= -{1\over
2} \gamma^5 \ , \qquad \quad P^a=im\gamma^a (1+s\gamma^5)
\equiv im\gamma^a P \ ,
\eqno(3.17)$$ where $s=\pm 1$. Consequently, a spinor field $\psi$
will transform according to
$$ \delta \psi = \alpha J \psi + \rho^a P_a \psi = -{\alpha \over 2}
\gamma^5 \psi +im \gam\rho P \psi \ \ , \eqno(3.18)$$
and the corresponding covariant derivative will be
$${\cal D}_\mu \psi = (\partial_\mu {\bf 1} + \omega_\mu J + e^a{}_\mu
P_a )\psi = \partial_\mu \psi -{1\over 2} \omega_\mu \gamma^5 \psi
+im\gam e_\mu P \psi \ \ . \eqno(3.19)$$

An $ISO(1,1)$ gauge invariant action  coupling fermions to the
model (2.19)
is then given by
$$S=\int d^2x \, \varepsilon^{\mu \nu} \left\{
\eta_A F^A{}_{\mu \nu} + \varepsilon_{ab} {\cal D}_\mu
q^a \left[ {\Lambda \over 2} {\cal D}_\nu q^b -{i\over 2} \alpha
\left( \bar \psi \Lambda^b {\cal D}_\nu\psi - {\rm H.C.} \right)
\right]\right\} \ \ , \eqno(3.20)$$
where H.C. denotes Hermitian conjugate, $\alpha$ is a dimensionless
coupling constant and
the matrix $\Lambda^b$ is given by
$$ \Lambda^b = \gamma^b +im[\gam q P , \gamma^b ] +2m^2 \gam q \gamma^b
\gam q P \ \ . \eqno(3.21)$$
The coefficient $-{i\over 2}$ in front of the spinorial part of the
action (3.20) has been chosen in such a way that it reproduces,
in the flat space-time limit, the
correct Dirac action for a Dirac spinor with mass $2m$, {\i.e.}
$$S_{\rm fermions} \longrightarrow \int d^2x \, \bar \psi [i\gam \partial
-2m]\psi \ \ . \eqno(3.22)$$

The relevant field equations arising from the action (3.20) are
$$\eqalignno{ {\delta S \over \delta \eta_A} =0 \quad  \rightarrow
\quad & F^A{}_{\mu \nu} =0 \ , &\hbox{(3.23a)} \cr
 {\delta S \over \delta \omega_\nu} =0 \quad \rightarrow
\quad & \partial_\mu \eta_2 + \varepsilon^a{}_b \eta_a e^b{}_\mu
-{\Lambda \over 2}
 q_a {\cal D}_\mu q^a &\cr &
+{i\alpha\over 4}\left[\bar \psi q_a \Lambda^a {\cal D}_\mu \psi
-{1\over 2}
\varepsilon_{ab}{\cal D}_\mu q^a\bar \psi \Lambda^b \gamma^5\psi
-{\rm H.C.} \right]
=0 \  , &\hbox{(3.23b)} \cr
{\delta S \over \delta e^a{}_\nu} =0 \quad \rightarrow
\quad & \partial_\mu \eta_a +\varepsilon_a{}^b \eta_b \omega_\mu
+{\Lambda\over 2} \varepsilon_{ab}{\cal D}_\mu q^b
&\cr & -{i\alpha \over 4} \left[\varepsilon_{ab}\bar \psi \Lambda^b
{\cal D}_\mu \psi +im \varepsilon_{bc} {\cal D}_\mu q^c \bar \psi
\Lambda^b \gamma_a P\psi - {\rm H.C.} \right]=0
 \  , &\hbox{(3.23c)} \cr
{\delta S \over \delta \bar \psi} =0 \quad \rightarrow
\quad & i \varepsilon^{\mu \nu}\varepsilon_{ab}
 {\cal D}_\mu q^a \left[
\Lambda^b {\cal
D}_\nu \psi + {1\over 2} {\cal D}_\nu \Lambda^b \psi \right] =0
 \ , &\hbox{(3.23d)} \cr }$$
where  ${\cal D}_\nu \Lambda^b$ denotes the covariant derivative
$$ {\cal D}_\nu \Lambda^b = im [{\cal D}_\nu \gam q P , \gamma^b] +
2m^2({\cal D}_\nu \gam q \gamma^b \gam q  P + \gam q \gamma^b {\cal
D}_\nu \gam q P ) \ \ . \eqno(3.24)$$
As in the previous example, the flat space-time condition Eq.(3.23a)
allows the gauge condition $q^a =0$, $\omega_\mu =0$, $e^a{}_\mu
=\delta^a{}_\mu$, so that Eq. (3.23d) becomes, as expected,
$$ [i \gam \partial -2m ]\psi =0 \ \  , \eqno(3.25)$$
whose plane wave solution  can be written as
$$ \psi^{(+)} ={e^{-ikx} \over \sqrt{E +2m}} \left(
\matrix{E+2m \cr -k_1 \cr} \right) \  , \qquad
\psi^{(-)} ={e^{ikx} \over \sqrt{E +2m}} \left(
\matrix{-k_1 \cr E+2m \cr} \right)\  , \eqno(3.26)$$
where the $\pm$ sign denotes the positive-negative energy solutions
(respectively)
and  $k_\mu = (E, k_1) = (\sqrt{(k_1)^2 + 4m^2} , k_1)$.
As an example, we shall consider a positive energy spinor $\psi^{(+)}$;
then Eqs.(3.23b-c) are equivalent to
$$ \eqalign{\partial_\mu \partial_\nu \eta_2 & = -\eta_{\mu \nu} \left[
{\Lambda \over 2} +4\alpha m^2 \right] +{\alpha}k_\mu k_\nu \cr
\eta_a & = \delta_a{}^\nu \varepsilon_\nu{}^\mu \partial_\mu \eta_2 \cr}
\eqno(3.27)$$
with solutions  given by
$$\eqalignno{\eta_2 &= -{1\over2} \left[{\Lambda \over 2} +4\alpha m^2
\right] x^2 +{\alpha \over 2} (kx)^2 +B\ \ , &\hbox{(3.28a)} \cr
\eta_a &= \delta_a{}^\nu \varepsilon_{\nu \mu} \left[{\alpha}
(kx) k^\mu -\left(4\alpha m^2 + {\Lambda \over 2}\right)x^\mu \right] \ \ ,
&\hbox{(3.28b)} \cr} $$
$B$ being an integration constant.

In this Section we have provided the main $ISO(1,1)$
 matter couplings to lineal gravity and solved the corresponding
equations of motion. At first glance, however, the gauge invariant
actions for the matter fields look rather complicated, due to the
explicit dependence on the Poincar\'e vectors $q^a$
which, in turn, is
necessary in order to have  Poincar\'e   gauge invariance
[see for instance Eqs. (3.10) and (3.20), (3.21)].
In the remaining part of this Section we shall show an elegant
and powerful procedure to obtain such gauge invariant actions.

Matter fields are supposed to belong to some suitable representation
 $(J, P_a)$ of the Poincar\'e algebra.
The zero momentum representation $(J, P_a \equiv 0)$, although trivial,
is indeed a representation of the Poincar\'e algebra. Moreover, gauge
invariant matter actions in the zero-momentum representation can be
quite easily found, as the matter fields transform
in this case only under the
Lorentz subgroup of the Poincar\'e group with generator $J$.

Given an arbitrary
representation $(J, P_a)$ for a generic matter field multiplet $\Phi$,
we can always find a corresponding multiplet $\tilde \Phi$ transforming
according to the zero-momentum representation $(J, 0$),
 the relation
between $\Phi$ and $\tilde \Phi$ being
$$ \tilde \Phi = ({\bf 1} - q^a P_a) \Phi \ \ . \eqno(3.29)$$
[By direct inspection, one can see that if $\delta \Phi =\alpha J \Phi +
\rho^a P_a \Phi$, then $\delta \tilde \Phi = \alpha J \tilde \Phi$.]
Consequently, Poincar\'e gauge invariant matter actions
in any arbitrary representation can be easily
found by first looking for a gauge invariant action
for matter fields $\tilde \Phi$ in the zero-momentum
representation, and successively by using Eq. (3.29) to express
$\tilde \Phi$ in terms of $\Phi$. As a specific example, we can
derive the Poincar\'e gauge invariant action for a spinor field of mass $2m$.
By choosing the representation (3.17) for the Poincar\'e algebra,
it turns out that the spinor
$$\chi = [{\bf 1} - im \gam q (1+s \gamma^5)]\psi
\equiv [{\bf 1} - im \gam q P]\psi  \ \ ,
\eqno(3.30)$$
transform according to
$$\delta \chi = \alpha J \chi = -{\alpha \over 2} \gamma^5
\chi \ \ , \eqno(3.31)$$
provided $\psi$ transforms according to Eq. (3.18).
 A gauge invariant action action for the $\chi$ spinor
is simply
$$S_F= - \int d^2x \, \varepsilon^{\mu \nu}
 \varepsilon_{ab} {\cal D}_\mu
q^a \left[{i\over 2} \left( \bar \chi \gamma^b {D}_\nu\chi - {\rm H.C.}
\right) -2m {\cal D}_\nu q^b \bar \chi \chi \right]
 \ \ , \eqno(3.32)$$
where $D_\mu \chi = \partial_\mu \chi + \omega_\mu J \chi$, as $\chi$
does not transform under translations.
By substituting Eq. (3.30) in Eq. (3.32) the action $S_F$ can be
written, after reshuffling, as
$$S=-{i\over 2}\int d^2x \, \varepsilon^{\mu \nu}
 \varepsilon_{ab} {\cal D}_\mu
q^a \left[  \bar \psi
\left( \gamma^b +im[\gam q P , \gamma^b ] +2m^2 \gam q \gamma^b
\gam q P
\right) {\cal D}_\nu\psi - {\rm H.C.} \right]
\ \ . \eqno(3.33)$$
Eq. (3.33)  is precisely the fermionic part of the action (3.20),
and is invariant under Poincar\'e gauge transformations (3.18)
with non trivial momentum generators (3.17).

\bigskip

\noindent{\bf IV.\quad DIMENSIONAL REDUCTIONS}
\medskip
\nobreak

In this Section we shall show that all the models previously discussed
as well as all the couplings with matter can be
obtained from suitable dimensional reductions of an $ISO(2,1)$ gauge
invariant theory in $2+1$ dimensions.
We shall denote
$2+1$ dimensional space-time indices
by $\alpha , \beta , \gamma , ... =0,1,2$, and the first two
components denote the corresponding $1+1$ dimensional space-time
indices $\mu , \nu , \rho ,...=0,1$.
In the same way internal $ISO(2,1)$ gauge indices will be labelled
 by  $A,B,C,...=0,1,2$, and the first  two values of $A, B, C, ...$ will
denote the corresponding $ISO(1,1)$ internal indices $a,b,c,...$, {\i.e.}
$A= (a,2)$ and so on.
 The convention on the
three dimensional  completely
antisymmetric symbol is $\varepsilon^{012}=1$.

In $2+1$ dimensions the Einstein Hilbert action
can be written as
$$S_{EH} ={1\over G} \int d^3 x \, \varepsilon^{\alpha \beta \gamma}
\eta_{AB} e^A{}_\alpha [\partial_\beta \omega^B{}_\gamma +
\varepsilon^B{}_{CD} \omega^C{}_\beta \omega^D{}_\gamma] \eqno(4.1)$$
($G$ being the Newton's constant with dimensions of a length) and
 $S_{EH}$ can be shown to be gauge invariant under the $ISO(2,1)$ gauge
transformations
$$\eqalign{\delta e^A{}_\alpha &= -\partial_\alpha \rho^A -\varepsilon^A{}_{BC}
e^B{}_\alpha \kappa^C - \varepsilon^A{}_{BC}\omega^B{}_\alpha \rho^C \cr
\delta \omega^A{}_\alpha &= -\partial_\alpha \kappa^A
-\varepsilon^A{}_{BC} \omega^B{}_\alpha \kappa^C \cr}\ \ . \eqno(4.2)$$
Here $\kappa^A$ and $\rho^A$ are  infinitesimal arbitrary functions
associated to Lorentz and translation transformations, respectively.

As is well known [2], introducing a suitable inner product for the
generators, $S_{EH}$ becomes the Chern Simons action of the $ISO(2,1)$
group.

That Eq.(4.2) are indeed the transformation laws for  $ISO(2,1)$
gauge potentials can be easily realized by redefining the  $SO(2,1)$
generators $J_A = -{1\over 2} \varepsilon_A{}^{BC} J_{BC}$ in  terms of the
usual Lorentz generators $J_{BC}$; then, the Poincar\'e  algebra written in
terms of$J_A$ and of the translation generators $P_A$  simply becomes
$$[J_A, J_B]=\varepsilon_{AB}{}^C J_C \ , \qquad
[J_A, P_B]=\varepsilon_{AB}{}^C P_C \ , \qquad
[P_A, P_B]=0\ . \eqno(4.3)$$

Introducing the Lie algebra valued gauge potential
$$A_\alpha = e^A{}_\alpha P_A + \omega^A{}_\alpha J_A \eqno(4.4)$$
and taking the algebra (4.3) into account,
 the transformations (4.2)
can be written in the form of the usual gauge transformations of a
non-Abelian gauge theory, namely
$$\eqalign{\delta A_\alpha & = - \Delta_\alpha u = -\partial_\alpha u
-[A_\alpha, u] \cr
u &= P_A \rho^A + J_A \kappa^A \cr} \eqno(4.5)$$

{}From the gauge potential $A_\alpha$
 one can construct the $ISO(2,1)$ field strength
$$\eqalign{& F_{\alpha \beta}  =[\Delta_\alpha , \Delta_\beta ]=
P_A T^A{}_{\alpha \beta} + J_B R^B{}_{\alpha \beta}
=P_A [\partial_\alpha e^A{}_\beta -\partial_\beta e^A{}_\alpha \cr &
+\varepsilon^A{}_{BC}(\omega^B{}_\alpha e^C{}_\beta +
\omega^C{}_\beta e^B{}_\alpha)] +
J_A[\partial_\alpha\omega^A{}_\beta - \partial_\beta\omega^A{}_\alpha
+\varepsilon^A{}_{BC}\omega^B{}_\alpha
\omega^C{}_\beta]\cr}\eqno(4.6)$$
However, as in the two dimensional case,
the components $e^A{}_\alpha$ of the gauge potential cannot be
interpreted as the physical {\it dreibein} without encountering
inconsistencies and, consequently,
 $T^A{}_{\alpha
\beta}$  cannot be identified with the physical
torsion [1, 12]. Again, we have to introduce a set of Poincar\'e coordinates
$q^A(x)$ transforming as Poincar\'e vectors
$$\delta q^A = \varepsilon^A{}_{BC} \kappa^B q^C +\rho^A \eqno(4.7)$$
and the physical {\it dreibein} $V^A{}_\alpha$ will be given by the covariant
derivative of the Poincar\'e vectors (for a detailed discussion on the
three dimensional case we refer the reader to Ref.[1])
$$ V^A{}_\alpha = {\cal D}_\alpha q^A = \partial_\alpha q^A
+\varepsilon^A{}_{BC} \omega^B{}_\alpha q^C + e^A{}_\alpha \ \ .\eqno(4.8)$$
Then, from Eq. (4.8) the physical torsion ${\cal T}^A{}_{\alpha \beta}$
will be given by
$$ {\cal T}^A{}_{\alpha \beta} = T^A{}_{\alpha \beta} +
\varepsilon^A{}_{BC} R^B{}_{\alpha \beta} q^C \ \ \eqno(4.9)$$
and, as expected, only in the physical gauge $q^A=0$, it coincides with $
T^A{}_{\alpha \beta}$.

Besides the couplings with matter fields and particles (see below), one can
construct other $ISO(2,1)$ gauge invariant actions  in terms of
gauge potentials and Poincar\'e coordinates. First of all there is the
cosmological constant term
$$ S_\Lambda = {\Lambda
\over 3!} \int d^3x\, \varepsilon^{\alpha \beta \gamma}
\varepsilon_{ABC} {\cal D}_\alpha q^A {\cal D}_\beta q^B
{\cal D}_\gamma q^C \ \ . \eqno(4.10)$$
Next, we can also consider the action invariant under the $SO(2,1)$
subgroup of the Poincar\'e group [13]
$$ S_E = {\gamma\over 4}\int d^3x \, \varepsilon^{\alpha \beta \gamma}
\eta_{AB} \omega^A{}_\alpha\left[\partial_\beta \omega^B{}_\gamma
+{2\over 3} \varepsilon^B{}_{CD} \omega^C{}_\beta
\omega^D{}_\gamma\right] \ \ , \eqno(4.11)$$
$\gamma$ being a dimensionless parameter.
However, if we add $S_E$ to the  Einstein Hilbert action $S_{EH}$,
the spin connection has to be taken as a function of the {\it dreibein}
rather than as an independent variable, so that we would actually have a
second order description. In order to have a first order formalism, together
with $S_E$ we shall also include a Lagrange multiplier term  enforcing the
(physical) torsionless condition [14,15] $$ S_\lambda = \int d^3x \,
\varepsilon^{\alpha \beta \gamma}\eta_{AB}  \lambda^A{}_\alpha
[\partial_\beta {\cal D}_\gamma q^B +
\varepsilon^{B}{}_{CD}\omega^C{}_\beta{\cal D}_\gamma q^D ] \ \ ,
\eqno(4.12)$$ where, in order to guarantee gauge invariance, the Lagrange
multiplier  fields $\lambda^A{}_\alpha$ transform according to $$\delta
\lambda^A{}_\alpha = \varepsilon^A{}_{BC} \lambda^B{}_\alpha \kappa^C \ \ .
\eqno(4.13)$$

Hence, the most general $ISO(2,1)$ gauge invariant action we can
consider is
$$S= S_{EH} + S_\Lambda + S_E + S_\lambda \ \ . \eqno(4.14)$$

The dimensional  reductions will  be always performed along the
$x^2$ direction, that will be compactified to a unit length. Moreover, to
confine the reduced theory on  a $1+1$ dimensional space-time, we shall set
$\partial_2 (anything) =0 $ by default. In this way all the $x^2$ integrals
in the $2+1$-dimensional actions will only give, after reduction,
 an overall unit factor.

First of all, we perform a dimensional reduction leading to the
Liouville gravity as $ISO(1,1)$ gauge theory.
To this purpose, we choose $\gamma =0$ and $\lambda^A{}_\alpha=0$.
This choice can  always be consistently performed as $\gamma$ is a
parameter and the gauge transformation for $\lambda^A{}_\alpha$ is
homogeneous (in other words, it is equivalent to  consider in $2+1$
dimensions only the Einstein Hilbert action plus the cosmological
constant term).

Then, we perform the dimensional reduction as shown in  Table A

\bigskip
\bigskip
\centerline{\it TABLE A}
\bigskip

\centerline{
\vbox{ \offinterlineskip \hrule
\def\tablerule{\noalign{\hrule}}
\halign{\vrule#&\strut\quad#\hfil\quad&
\vrule#&\quad\strut\hfil#\quad&\vrule#\cr
height4pt&\omit&&\omit&\cr
&\multispan3\hfil {\bf Dimensional Reduction A}\hfil &\cr
height4pt&\omit&&\omit&\cr
height4pt&\omit&&\omit&\cr \tablerule
height4pt&\omit&&\omit&\cr
&\hfil 2+1 Dimensions \hfil&& \hfil 1+1 Dimensions \hfil&\cr
height4pt&\omit&&\omit&\cr\tablerule
height4pt&\omit&&\omit&\cr
&\hfil $e^2{}_2$ \hfil&& \hfil $-\eta_2$ \hfil&\cr
height4pt&\omit&&\omit&\cr\tablerule
height4pt&\omit&&\omit&\cr
&\hfil $e^a{}_\mu$\hfil &&\hfil $e^a{}_\mu$\hfil&\cr
height4pt&\omit&&\omit&\cr\tablerule
height4pt&\omit&&\omit&\cr
&\hfil$\omega^2{}_\mu$ \hfil&& \hfil$\omega_\mu$\hfil&\cr
height4pt&\omit&&\omit&\cr\tablerule
height4pt&\omit&&\omit&\cr
&\hfil$\omega^b{}_2$\hfil && \hfil$\eta^b$\hfil&\cr
height4pt&\omit&&\omit&\cr\tablerule
height4pt&\omit&&\omit&\cr
&\hfil$\rho^a$\hfil && \hfil$\rho^a$\hfil&\cr
height4pt&\omit&&\omit&\cr\tablerule
height4pt&\omit&&\omit&\cr
&\hfil$\kappa^2$\hfil && \hfil$\alpha$\hfil&\cr
height4pt&\omit&&\omit&\cr\tablerule
height4pt&\omit&&\omit&\cr
&\hfil$q^a$\hfil && \hfil$q^a$\hfil&\cr
height4pt&\omit&&\omit&\cr\tablerule}}}
\bigskip
\bigskip
\noindent all the remaining $2+1$ dimensional quantities can be chosen
to vanish.
In particular, we set $\rho^2=0$ and $\kappa^a=0$; this means that we
are performing
(in the internal space) the same dimensional reduction induced by the
elimination of the corresponding dimension in the space-time:
in fact, according to the elimination of the $x^2$ axes in the
space-time, we retain only the generators of the translation
in the first two components ($\rho^a \ne 0$ and $\rho^2 =0$)
and the only possible Lorentz
transformation, {\it i.e.} a boost along the direction 1 in the internal space,
whose generator is  $J_2$ ($\kappa^2 \ne 0$ and $\kappa^a =0$).
In other words, if we denote by $\hat P_A$ and $\hat J_A$ the
$ISO(2,1)$ generators (we introduce the hat to avoid confusion with the
corresponding $ISO(1,1)$ generators), the $ISO(1,1)$ generators we get
from the dimensional reduction in Table A are $P_a =\hat P_a$ and
$J=\hat J_2$.

It is not difficult to check that the $ISO(2,1)$ gauge transformations
(4.2) and (4.7)  reproduce, with the proper identifications
specified in Table A, the correct
$ISO(1,1)$ gauge transformations (2,2a), (2.5)
and (2.16).  [Here and in the following, we shall denote this property
by {\it consistency} criterium; this means that a given dimensional
reduction not only  maps $2+1$ dimensional fields
 into $1+1$ dimensional fields, but also that the $2+1$ dimensional
gauge transformations are consistently mapped into the corresponding
$1+1$ dimensional transformations. This is the reason why we
dimensionally reduce not only the fields but also the gauge parameters
and, consequently, the generators.]

By substituting the content of Table A in the action $ S_{HE} +
S_\Lambda$ one gets
$$ S_{HE} +
S_\Lambda \rightarrow {1\over G}
\int d^2x \, \varepsilon^{\mu \nu} \bigl[
\eta_A F^A{}_{\mu \nu}  +{G\Lambda\over 2} (\eta_2 -\varepsilon^a{}_b\eta_a
q^b)
 \varepsilon_{ab} {\cal D}_\mu q^a {\cal D}_\nu q^b
\bigr] \ \ , \eqno(4.15)$$
which is, up to a constant $G$, the action leading to the Liouville
gravity described in Sec. II with cosmological constant $G\Lambda$
(see Eq. (2.22)). Notice that, in spite of the fact that
$\partial_2(anything)=0$ and  $q^2=0$, the dimensional reduction
maps ${\cal D}_2 q^2$ into the gauge invariant structure $\eta_2 -
\varepsilon^a{}_b \eta_a q^b$.

But there is another $ISO(2,1)$ dimensional reduction leading to the Liouville
gravity.
In fact, let us consider the action (4.14) with $\Lambda=0$ ({\it i.e.}
without cosmological constant term) and with $\lambda^A{}_\alpha =0$
(this choice is always possible for any dimensional reduction, the
transformation laws for $\lambda^A{}_\alpha$ being homogeneous) and
perform the reduction summarized in Table B

\bigskip
\bigskip
\centerline{\it TABLE B}
\bigskip

\centerline{
\vbox{ \offinterlineskip \hrule
\def\tablerule{\noalign{\hrule}}
\halign{\vrule#&\strut\quad#\hfil\quad&
\vrule#&\quad\strut\hfil#\quad&\vrule#\cr
height4pt&\omit&&\omit&\cr
&\multispan3\hfil {\bf Dimensional Reduction B}\hfil &\cr
height4pt&\omit&&\omit&\cr
height4pt&\omit&&\omit&\cr \tablerule
height4pt&\omit&&\omit&\cr
&\hfil 2+1 Dimensions \hfil&& \hfil 1+1 Dimensions \hfil&\cr
height4pt&\omit&&\omit&\cr\tablerule
height4pt&\omit&&\omit&\cr
&\hfil $e^A{}_\alpha$ \hfil&& \hfil $0$ \hfil&\cr
height4pt&\omit&&\omit&\cr\tablerule
height4pt&\omit&&\omit&\cr
&\hfil $\omega^2{}_\mu$\hfil &&\hfil $\omega_\mu$\hfil&\cr
height4pt&\omit&&\omit&\cr\tablerule
height4pt&\omit&&\omit&\cr
&\hfil$\omega^2{}_2$ \hfil&& \hfil$-\eta_2 \sqrt{M^2 /2}$\hfil&\cr
height4pt&\omit&&\omit&\cr\tablerule
height4pt&\omit&&\omit&\cr
&\hfil$\omega^b{}_2$\hfil && \hfil$\eta^b$\hfil&\cr
height4pt&\omit&&\omit&\cr\tablerule
height4pt&\omit&&\omit&\cr
&\hfil$\omega^a{}_\mu$\hfil && \hfil$e^a{}_\mu \sqrt{M^2/2}$\hfil&\cr
height4pt&\omit&&\omit&\cr\tablerule
height4pt&\omit&&\omit&\cr
&\hfil$\kappa^2$\hfil && \hfil$\alpha$\hfil&\cr
height4pt&\omit&&\omit&\cr\tablerule
height4pt&\omit&&\omit&\cr
&\hfil$\kappa^a$\hfil && \hfil$\rho^a \sqrt{M^2/2}$\hfil&\cr
height4pt&\omit&&\omit&\cr\tablerule
height4pt&\omit&&\omit&\cr
&\hfil$\rho^A$\hfil && \hfil$0$\hfil&\cr
height4pt&\omit&&\omit&\cr\tablerule
}}}
\bigskip
\bigskip
\noindent
where the constant $M\ne 0$ (with dimensions of a mass) has been introduced
in order to maintain the correct dimensions of the fields and
gauge parameters.

Some comments are in order. First of all we notice that since
$e^A{}_\alpha$ (as well as $\rho^A$) has been set equal to zero, this is a
 reduction from $ISO(2,1)$ to  an $SO(2,1)$-type group
 (and with this choice only the action $S_E$ survives in $S$).
Moreover, from Table B it is clear that the relation between the
two sets of generators (before and after the reduction)
 is $\hat J_a = P_a$ and $\hat J_2 =J$.

Using Table B in the transformations (4.2)
we obtain the following  gauge transformations for the reduced degrees
of freedom,
$$\eqalignno{ \delta\eta_2 & =\varepsilon_{ab} \eta^a \rho^b
&\hbox{(4.16a)} \cr
\delta \eta^a &= \alpha \varepsilon^a{}_b \eta^b + {M^2\over 2}
\eta_2\varepsilon^a{}_b \rho^b &\hbox{(4.16b)}\cr
\delta \omega_\mu &=-\partial_\mu \alpha - {M^2\over 2} \varepsilon_{ab}
e^a{}_\mu \rho^b &\hbox{(4.16c)}\cr
\delta e^a{}_\mu &= -\partial_\mu \rho^a
 +\alpha \varepsilon^a{}_b e^b{}_\mu -\varepsilon^a{}_b \omega_\mu
\rho^b &\hbox{(4.16d)}\cr}$$
{}From Eqs.(4.16) it is apparent that these are not $ISO(1,1)$ gauge
transformations (compare with Eqs. (2.5)). This is not surprising
because, as we already said,
this should be an $SO(2,1)$-type group. As a matter
of fact,
the transformations (4.15) can be rewritten as gauge transformations of
the $SO(2,1)$ de-Sitter group.
To this purpose, let us introduce the $SO(2,1)$ de-Sitter algebra
$$ [P_a , J]= \varepsilon_a{}^b P_b \ , \qquad [P_a, P_b]=
\varepsilon_{ab} J {M^2\over2} \ \ . \eqno(4.17)$$
Then, using the algebra (4.17),
the transformations  (4.16)
can be rewritten as
$$\eqalign{\delta A_\mu & = - \Delta_\mu u = -\partial_\mu u - [A_\mu , u]
\ \ , \cr
\delta \eta & = [ u , \eta] \ \ , \cr
A_\mu & = e^a{}_\mu P_a + \omega_\mu J \ \ , \cr
\eta & =\eta^a P_a + \eta_2 J \ \ , \cr
u & = \rho^a P_a + \alpha J \ \ . \cr
} \eqno(4.18)$$
Consequently, by construction
 the reduction $ISO (2,1) \rightarrow {\rm de-Sitter}\
SO(2,1)$ is consistent in the sense previously explained.  The field
strength corresponding to the gauge potential $A_\mu$ is
$$\eqalign{ F_{\mu \nu}& = [\Delta_\mu , \Delta_\nu ] =
 [\partial_\mu
e^a{}_\nu -\partial_\nu e^a{}_\mu
+ \varepsilon^a{}_b (\omega_\mu e^b{}_\nu - \omega_\nu e^b{}_\mu)]P_a
\cr &
+ \left(\partial_\mu \omega_\nu - \partial_\nu \omega_\mu + {M^2\over 2}
\varepsilon_{ab} e^a{}_\mu e^b{}_\nu \right) J \cr } \eqno(4.19)$$
and it transforms according to
$$ \delta F_{\mu \nu} = [u , F_{\mu \nu}] . \eqno(4.20)$$

The algebra (4.17) possesses an invariant, non degenerate bilinear form
given by the Killing metric
$$g_{AB} \equiv\left( \matrix{-{M^2\over 2} \eta_{ab} & 0\cr 0&1\cr}\right)
\eqno(4.21)$$
where now the indices $A,B$ run from $0$ to $2$. Using this metric
(4.21) and
the definition $T_A= (T_0, T_1, T_2)=(P_0, P_1, J)$ the algebra (4.17)
can be written in compact form
$$ [T_A, T_B]= f_{AB}{}^C T_C= \varepsilon_{ABD} g^{DC}T_C \ \
\eqno(4.22)$$

By taking into account Table B, Eqs. (4.19) and (4.21), the $2+1$
dimensional action $S$ is mapped, under reduction, into the following
de-Sitter $SO(2,1)$ gauge invariant action
$$ S \rightarrow \int d^2x \varepsilon^{\mu \nu} g_{AB} \eta^A
F^B{}_{\mu \nu} \ \ , \eqno(4.23)$$
that is precisely the de-Sitter $SO(2,1)$ model leading to the Liouville
gravity analyzed in Refs. [11].

The third dimensional reduction we shall consider is the one
 reproducing the
$ISO(1,1)$ black-hole model (2.19).  The reduction is summarized in Table C.

\bigskip
\bigskip
\centerline{\it TABLE C}
\bigskip

\centerline{
\vbox{ \offinterlineskip \hrule
\def\tablerule{\noalign{\hrule}}
\halign{\vrule#&\strut\quad#\hfil\quad&
\vrule#&\quad\strut\hfil#\quad&\vrule#\cr
height4pt&\omit&&\omit&\cr
&\multispan3\hfil {\bf Dimensional Reduction C}\hfil &\cr
height4pt&\omit&&\omit&\cr
height4pt&\omit&&\omit&\cr \tablerule
height4pt&\omit&&\omit&\cr
&\hfil 2+1 Dimensions \hfil&& \hfil 1+1 Dimensions \hfil&\cr
height4pt&\omit&&\omit&\cr\tablerule
height4pt&\omit&&\omit&\cr
&\hfil $e^a{}_\mu$ \hfil&& \hfil $e^a{}_\mu$ \hfil&\cr
height4pt&\omit&&\omit&\cr\tablerule
height4pt&\omit&&\omit&\cr
&\hfil $e^2{}_\mu$\hfil &&\hfil $G\omega_\mu$\hfil&\cr
height4pt&\omit&&\omit&\cr\tablerule
height4pt&\omit&&\omit&\cr
&\hfil$e^2{}_2$ \hfil&& \hfil$ \beta ={\rm const.}$\hfil&\cr
height4pt&\omit&&\omit&\cr\tablerule
height4pt&\omit&&\omit&\cr
&\hfil$\omega^a{}_\mu$\hfil && \hfil$0$\hfil&\cr
height4pt&\omit&&\omit&\cr\tablerule
height4pt&\omit&&\omit&\cr
&\hfil$\omega^2{}_\mu$\hfil && \hfil$\omega_\mu$\hfil&\cr
height4pt&\omit&&\omit&\cr\tablerule
height4pt&\omit&&\omit&\cr
&\hfil$\omega^a{}_2$\hfil && \hfil$0$\hfil&\cr
height4pt&\omit&&\omit&\cr\tablerule
height4pt&\omit&&\omit&\cr
&\hfil$\omega^2{}_2$\hfil && \hfil$0$\hfil&\cr
height4pt&\omit&&\omit&\cr\tablerule
height4pt&\omit&&\omit&\cr
&\hfil$\lambda^a{}_\mu$\hfil && \hfil$0$\hfil&\cr
height4pt&\omit&&\omit&\cr\tablerule
height4pt&\omit&&\omit&\cr
&\hfil$\lambda^2{}_\mu$\hfil && \hfil$0$\hfil&\cr
height4pt&\omit&&\omit&\cr\tablerule
height4pt&\omit&&\omit&\cr
&\hfil$\lambda^a{}_2$\hfil && \hfil$\eta^a /G$\hfil&\cr
height4pt&\omit&&\omit&\cr\tablerule
height4pt&\omit&&\omit&\cr
&\hfil$\lambda^2{}_2$\hfil && \hfil$(\eta_2 -\varepsilon_{ab} \eta^a
q^b)/G^2$\hfil&\cr
height4pt&\omit&&\omit&\cr\tablerule
height4pt&\omit&&\omit&\cr
&\hfil$q^a$\hfil && \hfil$q^a$\hfil&\cr
height4pt&\omit&&\omit&\cr\tablerule
height4pt&\omit&&\omit&\cr
&\hfil$\kappa^2$\hfil && \hfil$\alpha$\hfil&\cr
height4pt&\omit&&\omit&\cr\tablerule
height4pt&\omit&&\omit&\cr
&\hfil$\kappa^a$\hfil && \hfil$0$\hfil&\cr
height4pt&\omit&&\omit&\cr\tablerule
height4pt&\omit&&\omit&\cr
&\hfil$\rho^a$\hfil && \hfil$\rho^a$\hfil&\cr
height4pt&\omit&&\omit&\cr\tablerule
height4pt&\omit&&\omit&\cr
&\hfil$\rho^2$\hfil && \hfil$G\alpha$\hfil&\cr
height4pt&\omit&&\omit&\cr\tablerule
}}}
\bigskip
\bigskip
\noindent
Among the various reductions we proposed, this is the less trivial.
In fact, whereas the mapping of the translation generators is the same
as in the Reduction A ({\it i.e.} $\hat P_a =P_a$),
to the generator $J$ does not correspond a single $ISO(2,1)$ generator
but, rather, the combination $J =\hat J_2 + G\hat P_2$, as can be easily
deduced by the fact that there are two $ISO(2,1)$ gauge potentials
corresponding to the  same potential $\omega_\mu$ (or, equivalently,
there are two $ISO(2,1)$ infinitesimal gauge functions corresponding
to $\alpha$).
However, since the $\hat P_A$ generators commute,
 this identification is
permitted. As a matter of fact,
by substituting $J =\hat J_2 + G\hat P_2$, $\hat J_a=0$, $\hat P_a =
P_a$ in the $ISO(2,1)$ algebra for the generators ($\hat P_A , \hat
J_A$), we  get the $ISO(1,1)$ algebra (2.3) for the generators $(J, P_a)$ and,
consequently, this is a $ISO(2,1) \rightarrow ISO(1,1)$ reduction.

As can be checked by direct inspection,
 the $ISO(2,1)$ gauge transformations are mapped, through Table C, into the
corresponding $ISO(1,1)$ gauge transformations, {\it i.e.} the reduction
is consistent.

Moreover, substituting the content of Table C in the action (4.14) one
gets
$$S \rightarrow {1\over 2G}
 \int d^2x \, \varepsilon^{\mu \nu} \left[
\eta_A F^A{}_{\mu \nu} + {\Lambda G \beta} \varepsilon_{ab} {\cal D}_\mu
q^a {\cal D}_\nu q^b \right] \ \ , \eqno(4.24)$$
namely a $1+1$ dimensional action leading to a black-hole solution with mass
$2\Lambda G \beta$.
A final remark concerns the $q^2$  coordinate; we did not fix in Table C
any reduction for this degree of freedom
as it never appears in the reduced action (4.24) ({\it i.e.}
in the action $q^2$ is always multiplied by terms
that vanish when the reduction is applied). Consequently,
not only it corresponds to a completely decoupled degree of freedom but
it looses any possible dynamics
  when the reduction is performed.

Next, let us turn to the dimensional reductions for the matter
fields. From now on, we shall always consider the dimensional reduction
given in Table C for the gauge potentials and  Poincar\'e coordinates.

First of all, we obviously have to give the $ISO(2,1)$ gauge
invariant actions in $2+1$ dimensional space-time.

For the scalar
field we
follow in $2+1$ dimensions the prescriptions introduced in  $1+1$
dimensions. We then
introduce a multiplet $\Phi = (\varphi^A, m^2 \varphi)$ transforming
according to the vectorial representation of $ISO(2,1)$, that
 is four dimensional and gives
$$\eqalign{\delta \varphi^A &= \varepsilon^A{}_{BC}\kappa^B \varphi^C
+\rho^A m^2 \varphi \ \ , \cr
\delta\varphi & =0 . \cr} \eqno(4.25)$$
The covariant derivative of the multiplet $\Phi$ (transforming as the
field $\Phi$ itself)
is defined as
$$ \eqalign{{\cal D}_\alpha \varphi^A &=\partial_\alpha \varphi^A
+ \varepsilon^A{}_{BC} \omega^B{}_\alpha \varphi^C +m^2 e^A{}_\alpha \cr
{\cal D}_\alpha \varphi & =\partial_\alpha \varphi \cr}\eqno(4.26)$$
 An $ISO(2,1)$ gauge invariant action
written in terms of $\Phi = (\varphi^A, m^2 \varphi$) is then
$$ S_S = -{3\over 4} \int d^3 x \, \varepsilon^{\alpha \beta \gamma}
\varepsilon_{ABC} {\cal D}_\alpha q^A {\cal D}_\beta q^B [(\varphi
{\cal D}_\gamma \varphi^C -{\cal D}_\gamma \varphi \varphi^C)
+{\cal D}_\gamma q^C (\varphi^D -m^2 \varphi q^D)(\varphi_D -m^2 \varphi
q_D)] \eqno(4.27)$$
As it happened in the $1+1$ dimensional case, the variation of $S_S$
with respect to $\varphi^A$ provides a relation between $\varphi$ and
$\varphi^A$ that substituted in the equation of motion for $\varphi$
and assuming the invertibility of the {\it dreibein} leads to
the Klein--Gordon equation in curved space-time for a scalar field
$\varphi$ with mass $3m$. [We do not worry of the fact that in this way
we always get the Klein--Gordon equation for
a scalar field with mass $nm$ in a $n$ dimensional
space-time;
 in fact the mass $nm$ can be always rescaled to any desired
value (provided one suitably normalizes the momentum generators).
 Alternatively, since  the action $S_S$ is given by the sum of two terms
which are separately
gauge invariant,
 we can always put a suitable (numerical)
 coefficient in front of
the $(\varphi^D -m^2 \varphi q^D)(\varphi_D -m^2 \varphi
q_D)$ term to adjust the mass of $\varphi$ to $m$ in any dimension
and without spoiling gauge invariance.
Finally, a third possibility is to add the gauge invariant term
$c  m^2 \varphi^2\varepsilon^{\alpha \beta \gamma} \varepsilon_{ABC}
{\cal D}_\alpha q^A{\cal D}_\beta q^B{\cal D}_\gamma q^C$ to the scalar
Lagrangian. In any dimension,
the numerical constant $c$ can always be chosen in such a
way that the mass of $\varphi$ becomes $m$. ]

Before performing the dimensional reduction, we shall consider
scalar field configurations for which $\varphi^2 = m^2 \varphi q^2$.
This configuration is consistent with the
dimensional reduction C we are going to apply. In fact, after the
reduction is performed, $\varphi^2$ transforms exactly in the same way
as $m^2 \varphi q^2$. Consequently, the relation
$\varphi^2 = m^2 \varphi q^2$ can be seen as an $ISO(2,1)$ gauge choice
(before applying the reduction) that however
does not break the residual $ISO(1,1)$ gauge invariance that the Reduction C
entails.

The remaining components of the multiplet $\Phi$
are then dimensionally reduced according to

\bigskip
\bigskip
\centerline{\it TABLE D}
\bigskip

\centerline{
\vbox{ \offinterlineskip \hrule
\def\tablerule{\noalign{\hrule}}
\halign{\vrule#&\strut\quad#\hfil\quad&
\vrule#&\quad\strut\hfil#\quad&\vrule#\cr
height4pt&\omit&&\omit&\cr
&\multispan3\hfil {\bf  Reduction C (for scalars)}\hfil &\cr
height4pt&\omit&&\omit&\cr
height4pt&\omit&&\omit&\cr \tablerule
height4pt&\omit&&\omit&\cr
&\hfil 2+1 Dimensions \hfil&& \hfil 1+1 Dimensions \hfil&\cr
height4pt&\omit&&\omit&\cr\tablerule
height4pt&\omit&&\omit&\cr
&\hfil $\varphi^a$ \hfil&& \hfil $m^{1/2}\varphi^a$ \hfil&\cr
height4pt&\omit&&\omit&\cr\tablerule
height4pt&\omit&&\omit&\cr
&\hfil$\varphi$ \hfil&& \hfil$ m^{1/2} \varphi$\hfil&\cr
height4pt&\omit&&\omit&\cr\tablerule}}}

One can easily check that the reduction is consistent; moreover, by
substituting the contents of Tables C and D  in the action $S_S$ one gets
$$\eqalign{S_S \rightarrow &
 {3\over2} m\beta \int d^2x \, \varepsilon^{\mu \nu}
\varepsilon_{ab} {\cal D}_\mu q^a \left\{ { \over } \bigl(\varphi {\cal D}_\nu
\varphi^b -{\cal D}_\nu \varphi \varphi^b \bigr)\right.
\cr & \left. + {\cal D}_\nu q^b\left[
{3\over2} (\varphi_c - m^2 q_c \varphi)(\varphi^c - m^2 q^c \varphi)
+{m^2\over2} \varphi^2 \right]\right\}} \ \ , \eqno(4.28)$$
which is $ISO(1,1)$ gauge invariant.
Assuming the invertibility of the {\it zweibein} ${\cal D}_\mu q^a$,
 the action (4.28)
reproduces the Klein-Gordon equation in curved space-time for a scalar
field $\varphi$ with mass $3m$.

As it is clear from a comparison between Eqs.(3.10) and (4.28),
the Poincar\'e  gauge invariant
 action for a scalar field in $2+1$ dimensions
 is a straightforward generalization of the corresponding
action in $1+1$ dimensions. As we shall see in the sequel, this is no
longer true for the fermion fields.

The lowest dimensional
 representation of the Clifford algebra in $2+1$ dimensions
is $2\times 2$. As is well known it can be constructed in terms of Pauli
matrices  $ \vec{\sigma}$; we shall choose $\gamma^A = (\sigma^3,
i\sigma^2 , -i  \sigma^1)$. Then, fermions in $2+1$ dimensions are
represented by 2 components spinors $\psi$.

 However, contrary to what happens both in $1+1$ and in $3+1$
dimensions,
 a non trivial representation of the Poincar\'e
algebra cannot  be realized in terms of $\gamma$ matrices
characterizing the lowest dimensional representation of the Clifford
algebra: in fact to
define the momentum generators $P_A$ we need a matrix playing the role of
$\gamma^5$, which is not available in $2+1$ dimensions with the
given representation for the $\gamma$ matrices as
$i\gamma^0 \gamma^1 \gamma^2 = {\bf 1}$ . In other words, since  the
$\gamma^A$ form a representation of the $SO(2,1)$ algebra,
 it is impossible to find a
representation  of the $ISO(2,1)$
algebra  with non trivial momentum only in terms of $\gamma^A$.

Consequently, we have to consider a higher dimensional representation
of the Clifford and
 $ISO(2,1)$ algebra and, correspondingly, higher dimensional
spinors. The next available dimension is four, and the most convenient
representation for the $ISO(2,1)$ algebra is in terms of the four
dimensional Dirac matrices $\Gamma^A$ in the chiral representation, namely

$$\Gamma^0=\left(\matrix{0&-{\bf 1}\cr-{\bf 1}&0\cr}\right)\ , \quad
\Gamma^i=\left(\matrix{0&\sigma^i\cr-\sigma^i&0\cr}\right)\ , \quad
\Gamma^5=\left(\matrix{{\bf 1}&0\cr0&{\bf -1}\cr}\right)\ ,
\eqno(4.29)$$
where $i=1,2$ and we used capital $\Gamma$ in order to avoid confusion
with the previously defined $2\times 2$ $\gamma$ matrices.
Then, a representation of the Poincar\'e algebra is given by
$$ J_A = -{1\over 4}\varepsilon_{ABC}\Gamma^B \Gamma^C \ , \qquad
P_A = {i\over 3} m \Gamma_A ({\bf 1} + s \Gamma^5)
\equiv {i\over 3} m \Gamma_A  \Pi \eqno(4.30)$$
and we shall consider therefore a four dimensional spinor
$\Psi$ transforming according to
$$\delta \Psi = u\Psi =(J_A \kappa^A + P_A \rho^A)\Psi \ \ .\eqno(4.31)$$
The four component spinor $\Psi$ can be written in terms of a
two component spinor $\psi$ as
$$ \Psi = \left( \matrix{\psi \cr \sigma^3({\bf 1} -{2\over 3} i m
\gam q ) \psi \cr}\right) \ \ , \eqno(4.32)$$
where $\gam q = \gamma^A q_A$ with $\gamma^A$ the $2 \times 2$ Dirac
matrices previously defined, and the matrix
$\sigma^3({\bf 1} -{2\over 3} i m
\gam q )$ in the lower components of $\Psi$ has been introduced in order to
have consistency with the transformations (4.31).

Next, we turn to the construction of a gauge invariant action.
{}From the transformations properties (4.31) it follows that the
Dirac conjugate spinor $\bar \Psi = \Psi^\dagger \Gamma^0$ transform as
$$\delta \bar \Psi = -\bar \Psi u = -\bar \Psi (J_A \kappa^A + P_A \rho^A)
\ \ \eqno(4.33)$$
so that the bilinear $\bar \Psi \Psi$ is gauge invariant. Moreover,
if we define a covariant derivative of $\Psi$ transforming as $\Psi$
itself, any bilinear of the type $\bar \Psi {\cal D}_\alpha \Psi$
will be gauge invariant too. Such a covariant derivative is clearly
given by
$${\cal D}_\alpha \Psi = (\partial_\alpha {\bf 1} + \omega^A{}_\alpha
J_A + e^A{}_\alpha P_A ) \Psi \ \ . \eqno(4.34)$$
Then, a gauge invariant action is given by
$$ S_F = {i\over 4}\int d^3x \,
\varepsilon^{\alpha \beta \gamma} \varepsilon_{ABC}
{\cal D}_\beta q^B{\cal D}_\gamma q^C\left[\bar \Psi
\Lambda^A {\cal D}_\alpha \Psi - {\rm H.C.} \right] \ \ , \eqno(4.35)$$
where the matrix $\Lambda^A$ is given by
$$ \Lambda^A = \Gamma^A -i{m\over 3}[\gam q \Pi , \Gamma^A] +
{2\over 9} m^2 \gam q \Pi \Gamma^A \gam q \Pi \ \ . \eqno(4.36)$$

After lengthy but straightforward manipulations
it can be shown that the action (4.35), once the decomposition (4.32) is
taken into account and assuming the invertibility of the {\it dreibein}
${\cal D}_\alpha q^A$,
leads to the correct Dirac equation in curved space-time for a
two-component
spinor $\psi$ with mass $m$.
In particular, it should be remarked that
whereas $\Psi$ carries a non trivial representation of the Poincar\'e
group, the two component spinor $\psi$ transforms according to $SO(2,1)$
as can be easily checked by substituting Eq. (4.32) in (4.31).
This expected feature is a consequence of the fact that it does not
exist a representation of the Poincar\'e algebra in terms
of the  $2 \times 2$ Dirac matrices $\gamma^A$ besides the
one with trivial momentum operators $P_A=0$
and, consequently, $\psi$ can only transform according to $SO(2,1)$.
In the same way,  the Poincar\'e gauge invariant action (4.35)
when expressed in terms of $\psi$ through Eq. (4.32) is equivalent
to a  gauge invariant action  written in terms of
of $\gamma^A$ Dirac matrices and with the spinor $\psi$ belonging to the
zero momentum representation of the Poincar\'e group.

We shall apply  the dimensional reduction given in Table C
to the action (4.35), with the following
reduction for the fermions
$$ \psi \longrightarrow  \sqrt{m}[{\bf 1} -im q_a \gamma^a (1+s
\gamma^5)]\chi
\equiv \sqrt{m}[{\bf 1} -im q_a \gamma^a P]\chi \ \ , \eqno(4.37)$$
where $\chi$ denotes the corresponding (two components) spinor in $1+1$
dimensions and the ($\gamma^a , \gamma^5$) matrices in the r.h.s. are
those defined in Sect. III.

By substituting Table C, Eqs. (4.32) and (4.37) in the fermionic action
$S_F$  we get
$$ S_F \rightarrow -i {\beta m\over 2}
\int d^2x \, \varepsilon^{\mu \nu}
 \varepsilon_{ab} {\cal D}_\mu
q^a \left[
\left( \bar \chi \Lambda^b {\cal D}_\nu\chi - {\rm H.C.} \right)
-im{\cal D}_\nu q^b \bar \chi \chi
\right] \ \ , \eqno(4.38)$$
$$ \Lambda^b = \gamma^b +im[\gam q P , \gamma^b ] +2m^2 \gam q \gamma^b
\gam q P \ \ , $$
which is the correct $ISO(1,1)$ gauge invariant action in $1+1$
dimensions leading to the Dirac equation in curved space-time for a
spinor $\chi$ of mass $m$.

As a final comment we notice that, as it happens to the dimensional
reduction leading to the $1+1$ dimensional black-hole model Eq. (4.24),
the Poincar\'e coordinate $q^2$ completely disappears from the reduced
action.

\bigskip

\noindent{\bf V.\quad CONCLUSIONS}
\medskip
\nobreak

In a proposal of several years ago [4], it was suggested that
gravitational equations for lineal gravity could be expressed only in terms of
the scalar curvature $R$, the simplest object encoding
all the local geometric properties of $1+1$ dimensional space-time.
Since then, many models have been proposed and studied. However, two of
them have attracted a particular interest recently: the so called
Liouville-gravity, and the ``black-hole'' model.
These two models possess many peculiar and interesting features, both
from the mathematical and physical point of view. They
 have
been successfully used in
statistical mechanics (especially the Liouville-gravity). The black-hole
model, instead, represents the simplest theoretical laboratory in which  all
the  main characteristic features of the  black-hole geometry can be studied.
Consequently, it could be used to gain  interesting insights in
 a wide class of physical phenomena that would be difficult to treat in a four
dimensional context (such as the
quantum evolution of a black-hole, including back reaction [7]).

{}From the mathematical point of view, these models have  interesting
gauge theoretical properties. In Ref. [8, 9] it has been shown that the
black-hole model can be written as a gauge theory of the centrally
extended Poincar\'e group, {\i.e.} the Poincar\'e group with a central
extension in the momentum algebra, whereas Liouville gravity can be
formulated as an $SO(2,1)$ de-Sitter gauge theory [11].

In this paper, we have shown that both these models can be cast in the
framework of a Poincar\'e gauge theory, without a central extension,
providing therefore a gauge theoretical approach for the two
models alternative to the ones discussed in Refs. [6,8,9].

In our formulation
 the component $e^a{}_\mu$ of the gauge potential along
the translation generators are not identified with the physical {\it
zweibein}. This is given by the Poincar\'e covariant
derivative  of the coordinates $q^a$, ${\cal D}_\mu q^a$.

 Our method also provides the possibility of writing
 the gauge invariant
couplings with particles, scalar and fermion fields. We have also
solved the
corresponding equations of motion.

\smallskip
\noindent
In Sect. IV we have shown that all the $ISO(1,1)$ gauge invariant
models previously considered as well as all the
gauge invariant couplings to matter
can be obtained by suitable dimensional reductions  of an $ISO(2,1)$ gauge
invariant theory.

Concerning the dimensional reductions,
a different approach  was adopted in Ref. [10],
where it was shown that both the de-Sitter-Liouville
gravity [11] and the centrally extended Poincar\'e model [8] can be
obtained as dimensional reductions of a centrally extended
$ISO(2,1)$ $2+1$-dimensional theory.
However, the results given in Ref. [10]
 have to be considered  on a different ground respect to those
we have presented in Sect. IV, basically because the starting and ending
gauge groups (before and after the reduction) in [10]
 are different from the ones considered by us.
In particular, the coupling to matter is difficult to realize even
 in a centrally extended $ISO(2,1)$ gauge theory.
On the contrary,
we have shown [1,12] that these problems are absent
in an $ISO(2,1)$ gauge theory of gravity without central extension.

In conclusion, we think that the most convenient  group
to  formulate gravity coupled to matter as a gauge theory
 in any dimension is the Poincar\'e group (see also Refs. [1, 12] for the
three and four dimensions).

Several aspects of the results we have presented deserve a deeper analysis
and will be the subject of future investigations.
Among them, there are the quantization of the models coupled to matter and
the study of the physical properties of the new solutions found in Section
II, with particle, scalar and spinor sources. Moreover,
one could investigate what
corresponds in $2+1$-dimensions to the two dimensional
solutions presented in this paper.
Finally, another interesting problem could be the analysis and the
classification of new possible two dimensional models for gravity which admit
a gauge theoretical treatment
and a dimensional reduction from gauge invariant
$2+1$-dimensional actions.

\bigskip\bigskip
\noindent{\bf NOTE}
\medskip
\nobreak
After the completion of this work we received a paper by A. Ach\'ucarro [16] in
which a different dimensional reduction from the three dimensional
Einstein-Hilbert action with a non-vanishing cosmological constant, to the
black-hole action, is proposed.

\vfill
\eject

\noindent{\bf APPENDIX}
\medskip
\nobreak
\bigskip

In this Appendix we shall show some technical but important points
related to the pure gauge solutions. In particular, we shall
analyze in which case the physical {\it zweibein} assumes the unphysical
vanishing value.

In all the models considered in this paper, we always have the vanishing
field strength equation,
that are solved by pure gauge potentials.
As is well known, a particular type of pure gauge potential is the
vanishing one. In order to pursue the analogy with ordinary gauge
theories with compact gauge groups, $A_\mu=0$ should give absence of
interaction and consequently free matter actions, in other words, a flat
space-time.

This does not happen in the conventional approach to gravity as a gauge
theory, where the {\it zweibein} is chosen as the component of $A_\mu$
along the translation generators and, consequently, $A_\mu=0$
corresponds to the degenerate case of a vanishing space-time metric.
In our approach, instead, the gauge potential is not the{\it zweibein}
 and   one can show that $A_\mu =0$ leads to a {\it zweibein} that is
given everywhere by the derivative $\partial_\mu q^a$ (see Eq. (2.7))
which, being the $q^a$ Cartesian coordinates with Minkowskian metric,
simply implies a flat space-time.

With our method even in the physical gauge where $q^a=0$ and $V^a{}_\mu
=e^a{}_\mu$ one can still maintain the geometrical interpretation of the
{\it zweibein}. In fact in selecting the physical gauge we automatically
choose a particular value of $\rho^a$ and it is this constraint that, as
we shall see in the sequel, prevents $e^a{}_\mu$ to assume the vanishing
value. Actually it is not difficult to see that the vanishing field
strength condition together with $q^a=0$ gauge choice  imply $e^a{}_\mu
= \delta^a{}_\mu$ modulo a local Lorentz  transformation, that is
precisely what one would expect: in the flat space time limit the {\it
zweibein} can be chosen to be  $\delta^a{}_\mu$, and the Lorentz
transformation up to which $e^a{}_\mu$ is defined is related to the
residual $SO(1,1)$ gauge freedom that the physical gauge entails.
On the other hands, if the field strength vanishes, one can choose as a
particular solution the vanishing connection $e^a{}_\mu=\omega_\mu=0$
that, however, is a gauge choice different from the physical one and
moreover fixes completely the gauge arbitrariness. This gauge condition
implies that the Poincar\'e coordinates coincide (up to a constant
Minkowskian 2-vector) with the space-time coordinates, {\it i.e.}
$q^a=\delta^a{}_\mu x^\mu + c^a$ and the physical {\it zweibein} is
again $V^a{}_\mu = \partial_\mu q^a =\delta^a{}_\mu$. Hence, within our
formalism, the {\it zweibein} assumes its physical value when the
physical gauge or the vanishing gauge potential are picked. Let us
discuss these statements more quantitatively.
A pure gauge connection is of the form
$$ A_\mu = U^{-1} \partial_\mu U \ \ , \qquad \ \ U=U(\Lambda , \rho)
\in ISO(1,1)
\eqno(A.1)$$
An $ISO(1,1)$ gauge group element $U$ can be written,
in the vectorial representation of $ISO(1,1)$,  as
$$U=\left( \matrix{\Lambda^a{}_b & \rho^a \cr 0 & 1\cr} \right) \ \ ,
\eqno(A.2)$$
where $\Lambda^a{}_b = \delta^a{}_b \cosh \alpha -\varepsilon^a{}_b
\sinh \alpha$ is a finite Lorentz boost and $\rho^a$ a finite
translation.
With the representation (A.2)
 the group multiplication is simply replaced by the matrix
product. Then a pure gauge potential can be written as
$$A_\mu = U^{-1} \partial_\mu U = \left( \matrix{(\Lambda^{-1})^a{}_c
\partial_\mu \Lambda^c{}_b & (\Lambda^{-1})^a{}_c \rho^c \cr 0&0\cr}\right)
\ \ . \eqno(A.3)$$
Eq. (A.3) should be compared with the (Lie algebra valued) gauge potential
in the vectorial representation, namely
$$A_\mu = \omega_\mu J + e^a{}_\mu P_a =
\omega_\mu \left(\matrix{\varepsilon^a{}_b & 0\cr0&0\cr}\right)+
e^b{}_\mu \left(\matrix{0&\delta^a{}_b \cr0&0\cr}\right) \ \ ,
\eqno(A.4)$$
leading to the expected values for a pure gauge connection
$$\eqalign{\omega_\mu &= -\partial_\mu \alpha \cr e^a{}_\mu &=
(\Lambda^{-1})^a{}_b \partial_\mu \rho^b = \Lambda_b{}^a \partial_\mu
\rho^b \cr} \eqno(A.5)$$
The
corresponding physical {\it
zweibein} is, up to a Lorentz transformation, the total derivative of
the gauged transformed  Poincar\'e coordinates $q^a$ with
gauge group element $U=(\Lambda , \rho)$. In fact, substituting Eqs. (A.5)
in $V^a{}_\mu = {\cal D}_\mu q^a = \partial_\mu q^a +\varepsilon^a{}_b
\omega_\mu q^b  + e^a{}_\mu$ we get
$$ V^a{}_\mu = {\cal D}_\mu q^a = \Lambda_b{}^a \partial_\mu
(\Lambda^b{}_c q^c +\rho^b) = \Lambda_b{}^a \partial_\mu \hat q^a \ \ ,
\eqno(A.6)$$
where we denoted by $\hat q^a$ the $U$-gauge-transformed of the
Poincar\'e coordinates $q^a$.

The space time metric the becomes
$$ds^2 = V^a{}_\mu \eta_{ab} V^b{}_\nu dx^\mu dx^\nu =
\partial_\mu \hat q^a \eta_{ab} \partial_\nu \hat q^b dx^\mu dx^\nu
\equiv \eta_{ab} d\hat q^a d\hat q^b \ \ . \eqno(A.7)$$

Even if the gauge potentials in Eq. (A.5) are chosen to vanish the
metric is still given by (A.7) because $V^a{}_\mu =\partial_\mu q^a$,
namely, in the absence of interaction ($\hat A_\mu =0$) the metric
tensor is Minkowskian. Consequently the physical {\it zweibein}
corresponding to the zero connection gauge choice is, up to a Lorentz
transformation, a Kronecker delta.

In the physical gauge where $q^a=0$ we have that $V^a{}_\mu =e^a{}_\mu$.
But the physical gauge condition together with Eq.(A.5) impose $e^a{}_\mu$
to be of the form
$$e^a{}_\mu =-\Lambda_b{}^a \partial_\mu (\Lambda^b{}_c q^b) \ \ ,
\eqno(A.8)$$
and the metric is flat in terms of the coordinates $x^\mu = \delta^\mu_a
\Lambda^a{}_b q^b$.

Nevertheless, there exists (as it should be) a choice of pure gauge
potentials corresponding to the degenerate case of vanishing {\it
zweibein}. It is given by
$$\eqalign{ \omega^{ab}{}_\mu & =0 \cr e^a{}_\mu &=-\partial_\mu q^a \cr}
\eqno(A.9)$$ and corresponds to the vanishing of $\hat q^a$.

\vfill
\eject

\noindent{\bf REFERENCES}
\medskip
\nobreak
\bigskip

\item{[1]} G. Grignani and G. Nardelli, {\it Phys. Rev.} {\bf D45}, 2719
(1992), and references therein. See also W. Drechsler, {\it Ann. Inst. H.
Poincar\'e}, {\bf A XXXVII} 155 (1982).
 \medskip
\item{[2]} A. Ach\'ucarro and P. Townsend, {\it Phys. Lett.} {\bf B 180},
85 (1986); E. Witten, {\it Nucl. Phys.} {\bf B311}, 46 (1988).
\medskip
\item{[3]} For a review see R. Mann, in {\it Proceedings of the Fourth
Canadian Conference on General Relativity and Relativistic
Astrophysics}, to be published. See also A. H. Chamseddine, {\it Phys.
Lett.} {\bf B256}, (1991) 379; I. M. Lichtzier and S. D. Odintsov, {\it
Mod. Phys. Lett.} {\bf A6}, (1991) 1953.
\medskip
\item{[4]} C. Teitelboim, {\it Phys. Lett.} {\bf B126}, 41 (1983), and
in {\it Quantum Theory of Gravity}, S. Christensen, ed. (Adam Higler,
Bristol, 1984);
R. Jackiw in {\it Quantum Theory of Gravity}, S. Christensen, ed. (Adam Higler,
Bristol, 1984) and {\it Nucl. Phys.} {\bf B252}, 343 (1985).
\medskip
\item{[5]} E. Witten, {\it Phys. Rev.} {\bf D44}, 314 (1991).
\medskip
\item{[6]} H. Verlinde,	in {\it The Sixth Marcel Grossman Meeting on
General Relativity}, H. Sato, ed. (World Scientific, Singapore, 1992).
\medskip
\item{[7]} C. Callan, S. Giddings, A. Harvey and A. Strominger, {\it
Phys. Rev.} {\bf D45}, 1005 (1992).
\medskip
\item{[8]} D. Cangemi and R. Jackiw, {\it
Phys. Rev. Lett.}, {\bf 69}, 233 (1992).
\medskip
\item{[9]} R. Jackiw, in {\it Recent Problems in
Mathematical Physics}, to be published in {\it Theor. Math. Phys.}.
\medskip
\item{[10]} A different approach has been provided  by
D. Cangemi, MIT Preprint CTP\# 2124, July 1992 (unpublished).
The author obtains the two models as dimensional reduction of a
centrally extended $ISO(2,1)$ gauge theory.
\medskip
\item{[11]} T. Fukuyama and K. Kamimura, {\it Phys. Lett.} {\bf 160B},
259 (1985); K. Isler and C. Trugenberger, {\it Phys. Rev. Lett.} {\bf
63}, 834 (1989); A. Chamseddine and D. Wyler, {\it Phys. Lett.} {\bf
B228}, 75 (1989).
\medskip
\item{[12]} G. Grignani and G. Nardelli, {\it Phys. Lett.} {\bf B264}, 45
(1991); G. Grignani and G. Nardelli, {\it Nucl. Phys.} {\bf B370}, 491
(1992).
\medskip
\item{[13]} S. Deser, R. Jackiw and S. Templeton, {\it Ann. Phys. (N.Y.)}
 {\bf 140} 372 (1982).
\medskip
\item{[14]} S. Deser and X. Xiang, {\it Phys.Lett.}, {\bf B263}, 39 (1991).
 \medskip
\item{[15]} S. Carlip, {\it Nucl. Phys.}, {\bf B362}, 111 (1991).
\medskip
\item{[16]} A. Ach\'ucarro, Tufts University preprint, August (1992)
(unpublished).

\end